\documentclass[longauth]{aa} 
\usepackage[colorlinks=true,linkcolor=blue,citecolor=blue]{hyperref}

\usepackage{graphicx}
\usepackage{multirow}
\usepackage{tabularx}
\usepackage{longtable,tabu}
\usepackage{lscape}
\usepackage{subfigure}
\usepackage{longtable}
\usepackage{pdflscape}
\usepackage{placeins}
\usepackage{capt-of}

\usepackage{multirow}
\def\sun{\odot}
\bibpunct{(}{)}{;}{a}{}{,} 
\usepackage{multirow}
\usepackage{subfigure}

\newcommand{\planet}{WASP-76\,b\,\,}

\newcommand{\vsys}{$v_{\mathrm{sys}}$\xspace}
\newcommand{\vtell}{$V_{\mathrm{tell}}$\xspace}
\usepackage{xspace}
\newcommand{\vrest}{$V_{\mathrm{rest}}$\xspace}

\newcommand{\kms}{km s$^{-1}$\xspace}
\newcommand{\kp}{$K_{\mathrm{p}}$\xspace}
\usepackage{amssymb}
\usepackage{pifont}
\usepackage{url}
\usepackage{soul}
\usepackage{threeparttable}
\usepackage{lipsum,multicol}
\usepackage{natbib}
\usepackage[varg]{txfonts}

\def \mjup{$M_\mathrm{Jup}$}
\def \rjup{$R_\mathrm{Jup}$}

\usepackage{verbatim}
\usepackage{adjustbox}
\usepackage{xcolor}

\begin{document}
	
	\title{The GAPS Programme at TNG}
        \subtitle{LXIX.The Dayside of \planet revealed by GIANO-B, HARPS-N and ESPRESSO: Evidence for Three-Dimensional Atmospheric Effects \thanks{Based on observations made with the Italian Telescopio Nazionale Galileo (TNG) operated on the island of La Palma by the Fundacion Galileo Galilei of the INAF at the Spanish Observatorio Roque de los Muchachos of the IAC in the frame of the program Global Architecture of the Planetary Systems (GAPS).} }

\titlerunning{The Dayside of \planet with GIARPS and ESPRESSO}
	
	\authorrunning{Guilluy et al.}

	\author{G. Guilluy \inst{1} 
    \and 
P.~Giacobbe\inst{1} 
\and
M.~Brogi\inst{1,2} 
\and
F.~Borsa\inst{3}
\and
J.~P.~Wardenier\inst{4}
\and
F.~Amadori\inst{1,5}
\and
P.~E.~Cubillos\inst{1,6}
\and
M.~Basilicata\inst{1}
\and
A.~S.~Bonomo\inst{1}
\and
A.~Sozzetti\inst{1} 
\and
I.~Carleo\inst{1}
\and
T.~Azevedo~Silva\inst{7}
\and
A. Bignamini\inst{8}
\and
M.~Damasso\inst{1}
\and
C.~Di Maio\inst{9}
\and
A.~Ghedina\inst{10}
\and
M.~Lodi\inst{10}
\and
L.~Mancini\inst{1,11,12}
\and
F.~Manni\inst{1,11}
\and
G.~Micela\inst{9}
\and
V.~Nascimbeni\inst{13}
\and
D.~Nardiello\inst{5,13,14}
\and
L.~Pino\inst{7}
\and
M.~Rainer\inst{3}
\and
G.~Scandariato\inst{15}
}
	\institute{
		INAF -- Osservatorio Astrofisico di Torino, Via Osservatorio 20, 10025, Pino Torinese, Italy 
\and
Department of Physics, University of Turin, Via Pietro Giuria 1, I-10125 Torino, Italy 
\and
INAF -- Osservatorio Astronomico di Brera, Via E. Bianchi 46, 23807 Merate, Italy 
\and
Institut Trottier de Recherche sur les Exoplanètes, Université de Montréal, Montréal, Québec, H3T 1J4, Canada
\and
Dipartimento di Fisica e Astronomia ``Galileo Galilei'', Università di Padova, Vicolo dell'Osservatorio 3, IT-35122, Padova, Italy 
\and
Space Research Institute, Austrian Academy of Sciences, Schmiedlstrasse 6, 8042 Graz, Austria 
\and
INAF -- Osservatorio Astrofisico di Arcetri, Largo E. Fermi 5, 50125, Firenze, Italy 
\and
INAF – Osservatorio Astronomico di Trieste, via Tiepolo 11, 34143 Trieste 
\and
INAF -- Osservatorio Astronomico di Palermo, Piazza del Parlamento, 1, I-90134 Palermo, Italy 
\and
Fundación Galileo Galilei-INAF, Rambla José Ana Fernandez Pérez 7, 38712 Breña Baja, TF, Spain 
\and
Department of Physics, University of Rome ``Tor Vergata'', Via della Ricerca Scientifica 1, 00133, Roma, Italy 
\and
Max Planck Institute for Astronomy, Königstuhl 17, 69117, Heidelberg, Germany 
\and
INAF – Osservatorio Astronomico di Padova, Vicolo dell'Osservatorio 5, 35122, Padova, Italy 
\and
Centro di Ateneo di Studi e Attività Spaziali "G. Colombo" – Università degli Studi di Padova, Via Venezia 15, I-35131, Padova, Italy 
\and
INAF -- Osservatorio Astrofisico di Catania, Via S. Sofia 78, I-95123 Catania, Italy
		} 
	\date{Received date ; Accepted date }

	\abstract
{The study of the atmosphere of ultra-hot Jupiters (UHJs) with equilibrium temperature $\geq$2000~K provides valuable insights into atmospheric physics under such extreme conditions. }
{We aim to characterise the dayside thermal spectrum of the UHJ \planet and investigate its properties. We analysed data gathered with three high-resolution spectrographs, specifically two nights with simultaneous observations of HARPS-N and GIANO-B, and four nights of publicly available ESPRESSO optical spectra. We observed the planet's dayside covering orbital phases between quadratures (0.25 < $\phi$ < 0.75).}
{We performed a homogeneous analysis of the  GIANO-B, HARPS-N and ESPRESSO data and co-added the signal of thousands of planetary lines through cross-correlation with simulated spectra of the planetary atmosphere.} 
{We report the detection of CO in the dayside atmosphere of \planet\ with a signal-to-noise ratio (S/N) of 10.4 in the GIANO-B spectra. In addition, we detect Fe~I in both the HARPS-N and ESPRESSO datasets, with S/N of 3.5 and 6.2, respectively. A signal from Fe~I is also identified in one of the two GIANO-B observations, with a S/N of 4.0. 
Interestingly, a qualitatively similar pattern -- with a weaker detection in one epoch compared to the other -- is also observed in the two HARPS-N nights. The GIANO-B results are therefore consistent with those obtained with HARPS-N.
Finally, we compared our strongest detections of CO (GIANO-B) and Fe~I (ESPRESSO), with predictions from Global Circulation Models (GCMs). Both cross-correlation and likelihood analyses favour the GCM that includes atmospheric dynamics over a static (no-dynamics) model when applied to the ESPRESSO data.
This study adds to the growing body of literature employing GCMs to interpret high-resolution spectroscopic measurements of exoplanet atmospheres.
}
{}
{}

	\maketitle

\section{Introduction}

Over the past 20 years, the characterisation of exoplanetary atmospheres has established itself as a fundamental and rapidly evolving research area in the field of exoplanets.  
From an observational perspective, high-resolution spectroscopy (HRS) using ground-based observatories, with current resolving powers ranging from $R = 25,000$ to 140,000, has proven to be a powerful technique for studying and characterising exoplanetary atmospheres (see the reviews by \citealt{Birkby2018, Brogi2021} and references therein).  
This observational technique allows us to:  
$(i)$ resolve molecular bands and atomic lines into individual spectral lines, enabling a better identification of different chemical components;  
$(ii)$ exploit the Doppler shift of the planet’s spectrum to disentangle it from both the Earth’s atmospheric transmission and the host star’s spectrum, thus allowing to determine the planet’s atmospheric circulation \citep[e.g.,][] {ehrenreich2020, Seidel2020,Seidel2021,Pino2022,Seidel2025}.

\noindent{Using} HRS, various chemical species have been detected in exoplanetary atmospheres across both visible and near-infrared (nIR) bands \citep[e.g.,][]{Snellen2010,Brogi2012, Birkby2013, Wyttenbach2015, Hoeijmakers2019, Giacobbe2021, Pelletier2023}, and atmospheric dynamics, such as winds and global circulation patterns, have been observed \citep[e.g.,][]{Brogi2016, Louden2015, ehrenreich2020, Kesseli2021, Seidel2025}.

Simultaneously, theoretical models have become increasingly sophisticated, enabling the testing of more complex planetary scenarios. Global Circulation Models (GCMs), which are 3D numerical simulations of atmospheric dynamics, now provide critical tools for detailed atmospheric studies \citep[e.g.,][]{Showman2013, Rauscher2014}.
Originally developed for Earth's climate studies, Global Circulation Models (GCMs) have been adapted to simulate exoplanet atmospheres, particularly for tidally locked hot Jupiters orbiting close to their stars. These models solve the equations of fluid dynamics and radiative transfer, capturing 3D variations in temperature, winds, and chemical composition across the planet. By accounting for spatial and temporal variations -- such as day-night contrasts and equatorial jets -- GCMs offer a consistent physical framework to interpret observations like phase curves, emission, and transmission spectra. In particular, they reveal how atmospheric circulation influences the shape and position of spectral lines, providing a more realistic alternative to traditional 1D models for high-resolution spectroscopy \citep{Showman2010}.

Ultra-hot Jupiters (UHJs), which orbit extremely close to their host stars ($P_{\mathrm{orb}} \lesssim 3$ days for a main-sequence star), are exposed to intense stellar irradiation, resulting in equilibrium temperatures that can exceed 2000~K. These extreme conditions make UHJs ideal laboratories for studying atmospheric physics in regimes that are inaccessible within the Solar System.
There is growing evidence that UHJs exhibit atmospheric properties that differ significantly from those of cooler hot Jupiters. At temperatures above $\sim$2000~K, molecules start to thermally dissociate and atom to ionise \citep[e.g.,][]{Arcangeli2018, Parmentier2018}.
At these high temperatures, H$^-$ becomes a dominant source of opacity and significantly shapes the emergent spectrum \citep[e.g.,][]{Gandhi2020}. Furthermore, it has been shown that the hotter the planet's equilibrium temperature, the more likely it is to exhibit a thermal inversion in its atmosphere \citep[e.g.,][]{Baxter2020}. As a result, UHJs may transition from a non-inverted to an inverted atmospheric structure as temperatures increase. UHJs thus offer a unique opportunity to test and refine GCMs.

Several UHJs have been extensively investigated in the literature (e.g., see \citealt{Stangret2024} and references therein) through both transmission and emission spectroscopy, utilising spectrographs mounted on both ground-based high-resolution and space-borne low-resolution instruments.

\planet ($M_\mathrm{p}$=0.894~\mjup, $R_\mathrm{P}$=1.854~$R_\mathrm{Jup}$) is a short-period ($P=1.81$~d) UHJ orbiting a bright F7 main sequence star WASP-76\,A ($T_\mathrm{eff}$ = $6329\pm65$~K, $V=9.52$, $K=8.243$). The parameters for \planet and its host star can be found in Table~\ref{tab_parameters}.
\planet equilibrium temperature is approximately $\sim$$2228$~K \citep{ehrenreich2020}; however, due to its tidal locking, the temperature on the dayside can be higher than $\sim$$3000$~K \citep[e.g.,][]{Garhart2020,Wardenier2025}. It is known that thermal inversion layers exist on the dayside of this planet, such that temperature increases with altitude \citep[e.g.,][]{Edwards2020, Yan2023, Silva2024}.

\noindent{\planet is a reference UHJ that has been extensively studied. Numerous atomic, ionised, and molecular species have been identified in its atmosphere at HRS  \citep{Seidel2019, ehrenreich2020, Casasayas2021, Deibert2021, Kesseli2021, Landman2021,Tabernero2021,Wardenier2021, azevedosilva2022, Gandhi2022,Kawauchi2022,Kesseli2022,SanchezLopez2022,Savel2022,Deibert2023,Gandhi2023,Wardenier2023,Yan2023, Pelletier2023, maguire2024, masson2024, mansfield2024, Silva2024}. A comprehensive reference for all the research performed on this planet can be found in Table 2 of \citet{Silva2024}.

\noindent{However}, most of these studies have relied on 1D atmospheric models for comparison. Yet, \planet's atmosphere is inherently three-dimensional and, by relying only on 1D approaches, we risk overlooking critical signatures of this 3D nature.
As recently emphasised by \citet{Wardenier2025}, the 3D thermal structure of the planet exerts an even stronger influence on emission line profiles than in transmission. In transmission spectroscopy, line depths are determined by altitude differences between the line core and the continuum pressure levels. In emission, however, the line contrast depends directly on the difference in thermal flux between these atmospheric layers. If the temperature increases with altitude (as in a thermal inversion), this results in emission lines; otherwise, absorption lines are expected. Since emission spectra integrate flux from the entire planetary disk — sampling both hotter dayside and cooler nightside regions — the resulting line shapes reflect a range of vertical temperature structures that cannot easily be captured by 1D models \citep{Beltz2021}. Additionally, 3D GCMs predict that emission-line Doppler shifts vary with the orbital phase, potentially explaining observed offsets in \kp vs \vrest maps \citep{Wardenier2025}. The interplay between a tidally locked rotation and spatial temperature variations may also cause emission line strengths to change throughout the orbit. Such effects have already been reported in UHJ emission studies, where cross-correlation signals for species like Fe, CO exhibit unexpected significant velocity-changes between pre- and post-eclipse observations. 
For \planet, \citet{Silva2024} did not detect a \kp\ offset for Fe~I in the pre- and post-eclipse spectra obtained with ESPRESSO, but reported a blueshift along the \vsys\ axis of $-4.7 \pm 0.3$~km\,s$^{-1}$ (we further discuss this result in Sect.~\ref{results}). In contrast, \citet{Yan2023} reported negative \kp\ offsets for CO, although with large uncertainties (\kp\ = $191^{+31}_{-54}$~km\,s$^{-1}$).\\
Motivated by these findings, we collected new GIARPS data and analysed them (see Sect.~\ref{observations} for further details) in combination with publicly available ESPRESSO spectra. Our ultimate goal is to analyse the dayside of \planet through the detection of Fe~I and CO and to assess the impact of atmospheric dynamics on the observed high-resolution spectra, studying the three-dimensional structure of \planet’s atmosphere. In particular, CO and Fe~I emission lines have been previously employed to investigate temperature inversions in UHJs \citep[e.g.,][]{Pino2020, Yan2020, Borsa2022, Yan2022a}, offering valuable insights into the thermal structure of their atmospheres. Owing to its high thermal dissociation temperature, CO is especially well suited to trace the hottest regions of the dayside in these extreme environments. In particular, CO and Fe~I have been detected in the dayside of \planet by \citet{Yan2023} and \citet{Silva2024}, respectively.

The paper is organised as follows. We describe the observations in Sect.~\ref{observations}. We detail the data reduction procedures in Sect.~\ref{data_analysis}. We highlight our findings in Sect.~\ref{results}. We finally discuss our results and draw our conclusions in Sect.~\ref{conclusions}.

\begin{table*}
    \caption{Properties of the planet \planet\ and its stellar host.}
    \small
    \centering
    \begin{tabular}{lcc}
        \hline \hline
        Parameter & Value & Reference \\
        \hline
        \noalign{\smallskip}
        \textbf{Star} && \\
        \noalign{\smallskip}
        Effective temperature, $T_\mathrm{eff}$ (K)              & $6329\pm65$              & \citet{ehrenreich2020} \\
        Radius, $R_*$ ($R_\mathrm{\sun}$)                 & $1.756\pm0.071$            & \citet{ehrenreich2020} \\
        Mass, $M_*$ ($M_\mathrm{\sun}$)                 & $1.458\pm0.021$          & \citet{ehrenreich2020} \\
          \hline
        \noalign{\smallskip}
        \textbf{Planet} && \\
        \noalign{\smallskip}
        Radius, $R_\mathrm{p}$ (\rjup) & $1.854^{+0.077}_{-0.076}$   & \citet{ehrenreich2020} \\ 
        Mass, $M_\mathrm{p}$ (\mjup) & $0.894^{+0.014}_{-0.013}$   & \citet{ehrenreich2020} \\
        eccentricity, $e$                                 & 0                             & \citet{ehrenreich2020} \\
        Orbital period, $P$ (days)                          & $1.80988198^{+0.00000064}_{-0.00000056}$ & \citet{ehrenreich2020} \\
        Mid-transit time, $T_\mathrm{c}$ (BJD)              & $58080.626165^{+0.000418}_{-0.000367}$    & \citet{ehrenreich2020} \\
        RV semi-amplitude, \kp\ (\kms)                       & $196.52\pm0.94$           & \citet{ehrenreich2020} \\
        Equilibrium temperature, $T_\mathrm{eq}$ (K)               & $2228\pm122$              & \citet{ehrenreich2020} \\
         \hline
    \end{tabular}
     \label{tab_parameters}
\end{table*}

\begin{table}
    \label{obslog}
    \footnotesize
    \caption{Observations log.}
    \centering
    \resizebox{\linewidth}{!}{
    \begin{tabular}{c c c c c} 
        \hline \hline
        Night & Instrument & S/N & $t_\mathrm{exp}$ [s] & $N_\mathrm{obs}$ \\
        \hline
        2023-10-11 [N1] & GIANO-B & 18.42 - 27.57 - 31.40 & 200.0 & 90 \\
                        & HARPS-N & 46.57 - 61.85 - 70.18 & 900.0 & 25 \\
        2023-11-15 [N2] & GIANO-B & 17.56 - 30.55 - 38.30 & 200.0 & 102 \\
                        & HARPS-N & 31.03 - 69.76 - 83.88 & 900.0 & 29 \\
        \hline
    \end{tabular}
    }
    \tablefoot{From left to right, we report: the date at the start of the observing
    night; the instrument; the S/N (minimum-average-maximum); the exposure time per spectrum $t_\mathrm{exp}$; and the number of
    observed spectra $N_\mathrm{obs}$.}
\end{table}

\begin{figure}
\centering
\includegraphics[width=0.9\linewidth]{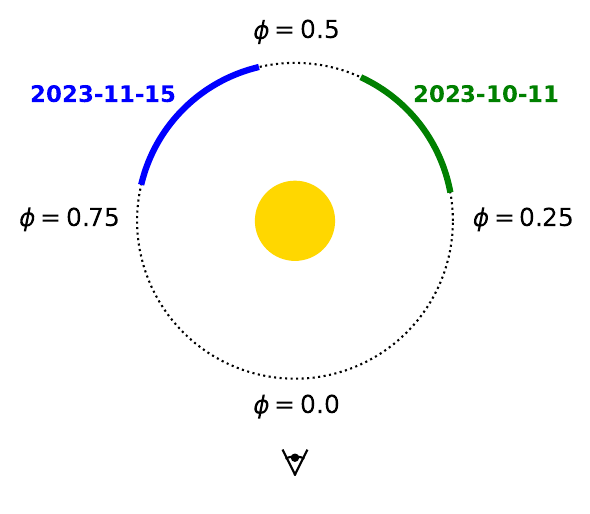}
\includegraphics[width=\linewidth]{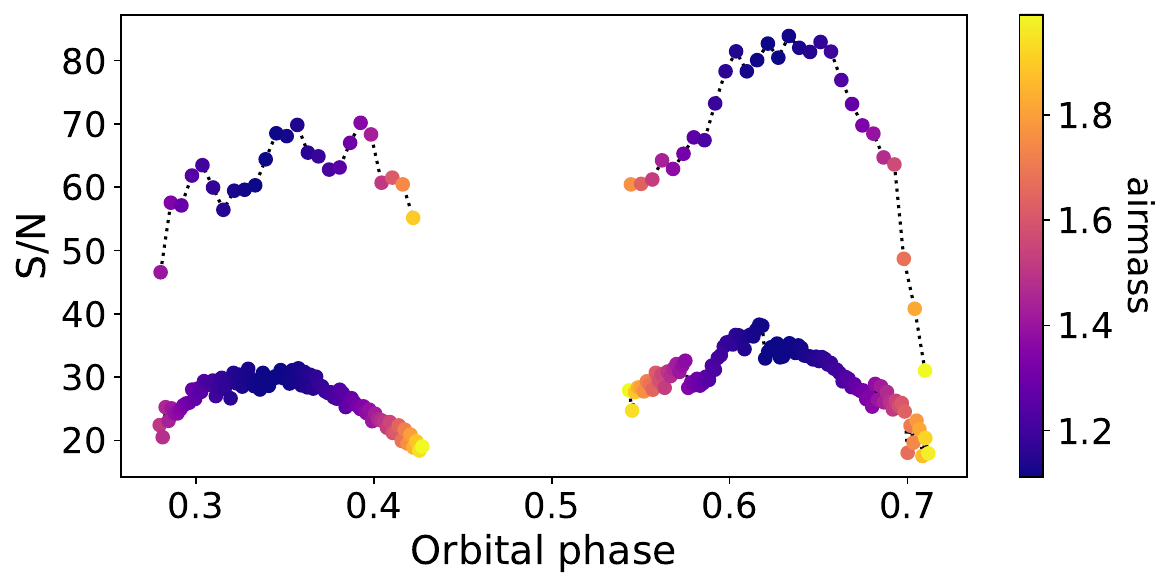}
\caption{ Top panel: Schematic representation of the orbit of WASP-76\,b, with the orbital phases covered in this study with GIARPS, colour-coded according to observing night. Bottom Panel: S/N averaged over orders as a function of the orbital phase for nights in emission, the markers' colour is proportional to the airmass. The exposures at a higher S/N ratio are the HARPS-N ones, while those at a lower S/N are from GIANO-B.}
\label{S/Nobit}
\end{figure}

\section{Observations}\label{observations}
We analysed two day-side emission observations of WASP-76 b (i.e.
2023-10-11 and 2023-11-15; hereafter N1 and N2) gathered with the
GIARPS observing mode of the Telescopio Nazionale Galileo (TNG, \citealt{GIARPS_claudi}) as part of the BRIDGES programme (PI F.~Borsa) within the GAPS (Global Architecture of Planetary System) Collaboration \citep{Covino2013}. We also re-analysed four ESPRESSO datasets (i.e. 2021-09-02, 2021-09-09, 2022-10-14, 2022-10-18, hereafter N3, N4, N5, N6), which we downloaded from the ESO public archive. The data were obtained as part of the programmes 1104.C-0350(U) and 110.24CD.004 of the Guaranteed Time Observation.

In this section, we describe the GIARPS observations while we refer the reader to \citet{Silva2024} for further details on the ESPRESSO dataset. 

 \noindent{When} observing in GIARPS mode, the TNG is capable of simultaneously acquiring high-resolution spectra in the optical range (0.39-0.69~$\mu$m) and nIR range (0.95-2.45~$\mu$m) using the HARPS-N ($R \approx 115,000$) and GIANO-B ($R \approx 50,000$) spectrographs.
For the GIANO-B observations, we employed an ABAB nodding pattern, which allows for optimal subtraction of thermal background noise and telluric emission lines.  GIANO-B covers four spectral bands in the nIR ($Y$, $J$, $H$, $K$), divided into 50 orders.
A detailed log of the GIARPS observations is provided in Table~\ref{obslog}.
Figure~\ref{S/Nobit} displays the average signal-to-noise ratio (S/N) in the GIANO-B spectra as a function of the planet's orbital phase and airmass for each night considered.  

The HARPS-N observations were carried out with the fibre B was put on the sky. 
HARPS-N covers 69 different orders for each fibre, utilising the echelle spectrograph design.

\section{Data Analysis} \label{data_analysis}
In this work, we fully leveraged the capabilities of the GIARPS instrument by analysing data from both GIANO-B and HARPS-N, and complemented this dataset with archival observations from ESPRESSO. Our goal was to maintain a consistent approach across the different datasets, focusing on the detection of CO and Fe~I in the dayside of WASP-76\,b.

\subsection{Data reduction and calibration} \label{reduction}

We processed the raw GIANO-B spectra using the GOFIO pipeline \citep{Rainer2018}, which handles flat-fielding, bad pixel correction, background subtraction, optimal extraction of the 1D spectra, and an initial wavelength calibration based on a Uranium-Neon lamp. To account for small temporal variations in the wavelength solution -- arising from the fact that it is typically derived at the end of the night and may be affected by instrumental instabilities occurring during the observing hours \citep{Giacobbe2021} -- we corrected and refined the wavelength solution. Specifically, we aligned all the spectra from a given night to a common reference frame by cross-correlating each exposure with the time-averaged observed spectrum of the target, which served as a template. We then exploited the stationarity of telluric lines to refine the wavelength calibration. In particular, after aligning the spectra, we performed a more accurate wavelength calibration by matching a set of telluric features in the time-averaged spectrum to a high-resolution transmission spectrum of Earth’s atmosphere, generated with the ESO Sky Model Calculator\footnote{\url{https://www.eso.org/observing/etc/bin/gen/form?INS.MODE=swspectr+INS.NAME=SKYCALC}}. However, due to the presence of spectral regions with either too many or too few telluric lines, it was not possible to apply this calibration refinement to all spectral orders \citep[e.g.,][]{Brogi_2018_Giano, Guilluy2022, Carleo2022,Basilicata2024}. We therefore excluded a set of spectral orders heavily affected by telluric contamination (orders 8–10 and 23–25), the Y-band region (orders 40–49) where GIANO-B exhibits a significant drop in throughput, and approximately ten additional orders (depending on the night) due to high residual drift or failures in the refined wavelength calibration procedure. 

For the HARPS-N observations, we calibrated the data using the standard Data Reduction Software (DRS v3.7.1; \citealt{Cosentino2012}) and extracted the ED2S files. The ESPRESSO spectra were reduced with the dedicated pipeline (DRS v3.2.5) provided by ESO and the ESPRESSO consortium \citep{Pepe2021}. For our analysis, we used the S2D spectra, which are referenced to the barycentric rest frame of the Solar System. In contrast, the GIARPS spectra are provided in the telluric rest frame.

For HARPS-N, we converted the wavelength calibration from air to vacuum, using the values retrieved from the file headers, to ensure consistency with the wavelength solutions of GIANO-B and ESPRESSO. No further realignment or wavelength correction was applied to either HARPS-N or ESPRESSO data, given the high intrinsic stability of both spectrographs. 

\subsection{Telluric and stellar removal}
We removed the telluric contamination by employing principal component analysis (PCA, \citealt{Giacobbe2021}) with the spectra aligned in the Earth's rest frame. Before applying PCA, we applied some preliminary steps \citep{Basilicata2024}. More precisely, once we converted the spectra into logarithmic space, we corrected for baseline flux variations by normalising each spectrum to its median value. We then identified and masked spectral channels contaminated by strong or saturated telluric lines with the same procedure adopted in \citet{Basilicata2024}.  Following, we computed the standard deviation of each spectral channel and its median value $\sigma_\mathrm{m}$ (after subtracting for each pixel the mean flux across the time axis, i.e., over all images) and masked pixels with variability exceeding a threshold -- defined as 1.5$\times\sigma_\mathrm{m}$ \citet{Basilicata2024}. The positions of the masked pixels were recorded and excluded from further analysis. This step ensured that highly variable or defective pixels did not bias the PCA decomposition, improving the robustness of the extracted components. 
After subtracting the time-averaged spectrum from all the spectra, we subtracted from each spectrum its mean value computed on the different spectral channels.
The PCA was computed from each spectral order. We used the python $pydl.pcomp$ routine to compute the eigenvectors (i.e. the principal components) and eigenvalues, and we built the matrix that should mainly describe the telluric contaminations via a linear combination of the principal components. This matrix was then subtracted from the original data, yielding a residual matrix where the dominant systematic trends were removed. For each dataset, we selected the optimal number of components to remove ($N_\mathrm{opt}$, see Table~\ref{pca_comp}), based on the standard deviation of the residuals ($\sigma$). Following an approach similar to that proposed by \citet{Basilicata2024} and references therein, we chose the number of components when the ratio $(\sigma_{i-1}-\sigma_i)/\sigma_{i-1}$ reached a plateau, where $i$ is the number of the PCA iteration.
We finally applied a high-pass filter to each row of the residual matrix to remove any possible residual correlation between different spectral channels \citet{Basilicata2024}.

\subsection{Species detection via cross-correlation} \label{cc}
\begin{figure}
	\centering
	\includegraphics[width=\linewidth]{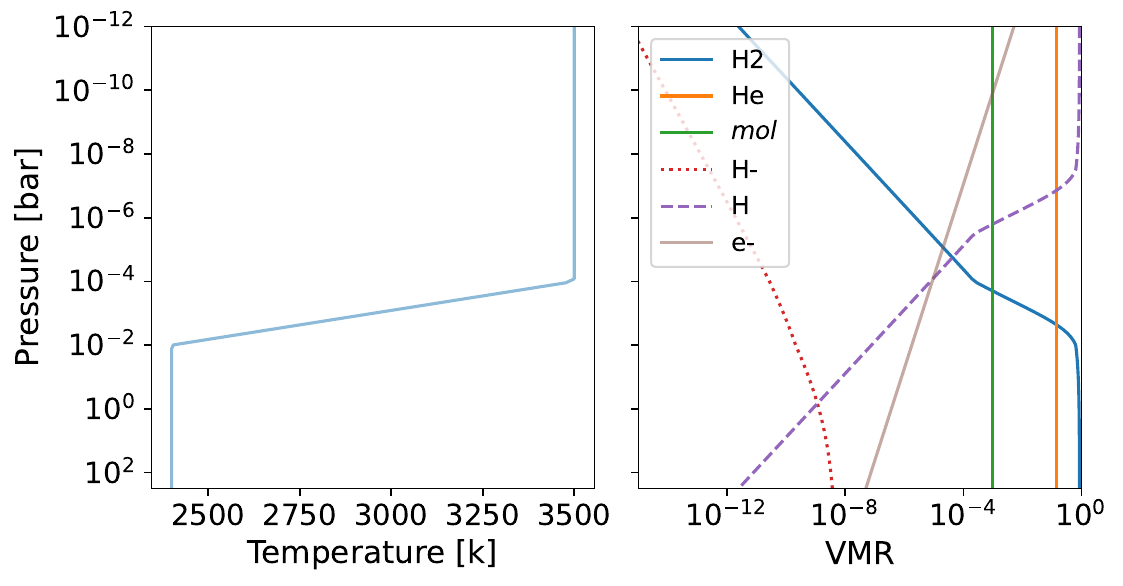}
\caption{Left Panel: Temperature-Pressure (TP) profile used for our cross-correlation analysis, from \citet{Yan2023}. Right panel: VMRs adapted in the single-species models used in this paper. The string "mol" represents each of the investigated molecules. }
	\label{TP_profile}
\end{figure} 
After the removal of telluric and stellar signals, we are left with residuals that contain both noise and the planetary signal. However, individual planetary lines have low S/N and remain buried beneath the noise level. To assess the presence of CO and Fe~I in our GIANO-B, HARPS-N, and ESPRESSO data, we therefore need to coherently combine the signal from all planetary lines. This can be achieved through the cross-correlation (CC) technique \citep[e.g.,]{Snellen2010, Brogi2012, Birkby2013}. For the initial signal recovery, we used 1D planetary emission models ($F_\mathrm{P}$) as cross-correlation templates, generated with the petitRADTRANS (pRT) code, version 3 \citep{Molliere2019}.
We used a pressure grid of 120 layers, equidistant in log-pressure $(-12 \leq \log_{10}[P \, (\text{bar})] \leq 2.5)$. For the spectral line lists, we employed the HITEMP\footnote{\url{https://hitran.org/hitemp/}} database for CO \citep{CO} and the KURUCZ line list\footnote{\url{http://kurucz.harvard.edu/}} for Fe~I.  We considered continuum contribution from collision-induced absorption (CIA) cross-sections for H\(_2\)-H\(_2\) \citep{Borysow2001, Borysow2002} and H\(_2\)-He pairs \citep{Borysow1988, Borysow1989, BorysowFrommhold1989}.
We additionally include H${^-}$ bound-free/free-free continuum opacity, since at the high temperatures of UHJs, H${^-}$ may be present in the atmosphere \citep[e.g.,][]{Arcangeli2018, Parmentier2018, Gandhi2020}. 
We followed the methodology proposed by \citet{Brogi2014} to construct the TP profile. We employed a two-point parametrisation, namely \((T_1, P_1)\) and \((T_2, P_2)\), where subscripts 1 and 2 indicate the bottom and the top of the stratosphere, respectively. For pressures higher than \(P_1\) or lower than \(P_2\), the temperatures are assumed to be isothermal; for pressures between \(P_1\) and \(P_2\), the temperature changes linearly with \(\log_{10}(P)\) with a gradient following the formula:\\
\begin{equation}
    T_{\mathrm{slope}} = \frac{T_1 - T_2}{\log P_1 - \log P_2},
\end{equation}

To determine the \(T_1, T_2, P_1, P_2\) values, we followed the prescription of \citet{Yan2023}. Specifically, we adopted an inverted TP profile with an intermediate approach between the HELIOS \citep{Malik2017} TP profiles calculated in the case of no heat redistribution and of full heat redistribution from the dayside to the nightside. Consequently, our TP profile was characterized by \( P_1 = 10^{-2} \,\mathrm{bar} \), \( P_2 = 10^{-4} \,\mathrm{bar} \), \( T_1 = 2400.0 \,\mathrm{K} \), and \( T_2 = 3500.0 \,\mathrm{K} \). The resulting profile is shown in the left panel of Fig.~\ref{TP_profile}.

The model assumes constant-with-altitude abundance (volume mixing ratio, VMR) profiles for the investigated chemical species\footnote{We adopted constant-with-altitude VMRs, as ionised CO and Fe~I are disfavoured in the atmosphere of WASP-76\,b. This is because (i) higher temperatures are predicted to dissociate CO's strong triple bond, and (ii) \citet{Silva2024} reported a non-detection of Fe~II in this planet's atmosphere.}, while for molecular hydrogen, atomic hydrogen, and ionic hydrogen, we changed the abundances as a function of the pressure following the formalism proposed by \citet{Parmentier2018}, see right panel of Fig.~\ref{TP_profile}.  To self-consistently compute the electron VMR profile, we would need to account for the ionisation of various species that are not included in the current model (e.g., alkali and other metals). Since such a detailed calculation lies beyond the scope of this study, we once again followed the simplified parametrisation of \citet{Parmentier2018}, adopting an increasing electron VMR with decreasing pressure, with a slope of approximately -0.4 in logarithmic space.

We scaled the planet model spectrum $F_\mathrm{P}$ to units of stellar continuum flux (with a black body at $T_\mathrm{eff}$, see Table~\ref{tab_parameters}) via equation (4) in \citet{Brogi2014}. The models were then convolved with the spectrograph instrument profile, corresponding to a Gaussian profile with an FWHM of 6~\kms,  2.6~\kms and 2.17~\kms for GIANO-B, HARPS-N and ESPRESSO, respectively. The lines in the spectrum of a rotating exoplanet are broadened because light from the receding limb is redshifted, and light from the approaching limb is blue-shifted. We thus also applied a kernel for planetary rotation broadening; we used a recipe similar to that of \citet{Gray} for the stellar broadening with a bell-shaped profile. 
For each night, we performed CC over all available spectral orders that were not discarded during the basic reduction steps \citep{Giacobbe2021} --except for CO, for which we considered only the K-band. We followed the CC methodology described in \citet{Gibson2020}, \citet{Cont2022}, and \citet{Nortmann2025}  to account for the uncertainties in the matrix of residual spectra (i.e., the data after PCA correction). For HARPS-N, we decided to exclude from our analysis the first spectrum of N1 and the last three spectra of N2, as they exhibited a significant drop in the S/N ratio (see Fig.~\ref{S/Nobit}). Conversely, all spectra from both ESPRESSO and GIANO-B were retained in the analysis.
}
The CC was computed at each orbital phase over a lag vector corresponding to planetary radial velocities (RVs) in the range $-300 \leq \mathrm{RV} \leq 300$~\kms, using velocity steps of 3~\kms, 1.5~\kms, and 1~\kms for GIANO-B, HARPS-N, and ESPRESSO, respectively.

 \noindent{The stellar} contamination represented one big challenge in our analysis. \planet\ is an ultra-hot Jupiter orbiting a hot F7-type star (T\(_{\text{eff}}\)=6329 K). However, the primary star has a candidate visual companion \citep{Wollert2015}. With a radius of $0.795\pm0.055$~$R_\mathrm{\sun}$ and an effective temperature of $4850\pm150$~K \citep{Fu2021}, WASP-76\,B is likely a late-G or early K-type dwarf \citep{ehrenreich2020}. This companion is separated by $\sim$0.44~arcsec \citep{Ginski2016, Ngo_2016, Bohn2020}, which means that our observations may be contaminated by light from the companion star and its possible activity, as the separation between the two stars is smaller than the slit width of GIANO-B and the fibre diameter of both HARPS-N and ESPRESSO. We attempted to correct the GIANO-B spectra using a similar method proposed by \citet{Flowers2019} and \citet{Chiavassa2019}. However, because the observed stellar spectrum is a combination of both stars, and we can not precisely quantify the level of contamination from the secondary star within our observations, especially considering that the contamination may vary during exposures, we were unable to model and remove the stellar spectrum from our data.  To mitigate stellar contamination, we adopted an alternative approach. 
Specifically, for each spectral order, we computed a 2D cross-correlation matrix, $CC2D$, with dimensions RV $\times$ $N_\mathrm{obs}$ (with $N_\mathrm{obs}$ the number of observations). We then defined a reference region around the stellar residuals (specifically, RV between $-20$ and $+20$~km/s with respect to the star’s radial velocity). Within this interval, we averaged the corresponding columns of the $CC2D$ matrix to produce a reference column vector. This vector was then fitted as a function of orbital phase using a second-order polynomial, hereafter referred to as $pccf$.
To account for amplitude variations among the individual columns of the $CC2D$ matrix, we assumed that the shape of the polynomial $pccf$ is preserved across RVs and allowed only for a linear rescaling.
For each column $i_\mathrm{RV}$ of the $CC2D$ matrix, we solved the following equation:
\begin{equation}
pccf_{i_\mathrm{RV}} = a \cdot pccf + c,
\end{equation}
where the coefficients $a$ and $c$ were determined using Singular Value Decomposition (SVD)\footnote{We emphasise that, in our case, SVD is not used to decompose the signal -- as is commonly done in the field -- but rather as a numerically robust method to solve a linear system. SVD is particularly suitable for solving systems involving matrices that are singular or nearly singular, as it provides improved numerical stability compared to the other numerical equations \citep{press2007numerical}.}.
Finally, the scaled polynomial $pccf_{i_\mathrm{RV}}$ was subtracted from each corresponding column of the $CC2D$ matrix.\footnote{We verified through injection-recovery tests that this filtering process does not alter the planetary signal. Additionally, since the planetary signal varies significantly with time in the telluric/barycentric rest frame, the filter --being applied column-- affects only one temporal element per RV, and thus its impact on the planetary signal is expected to be negligible.}
We then combined the $CC2D$ matrices across the orders (see the grey panels for each night in appendix Figs.~\ref{CO_Fe}, and \ref{Fe_espresso}), on nights, and on orbital phases, after shifting them into the planet's rest frame, assuming a circular orbit \citep[e.g.,][]{Brogi_2018_Giano, Giacobbe2021, Guilluy2022} and accounting for the planet’s time-dependent RV and the systemic velocity (\vsys). As we found discrepancies in the systemic velocity values reported in the literature \citep[e.g.,][]{West2016, Sobiran2018,ehrenreich2020,Silva2024}, we adopted, for each instrument, a value of \vsys equal to the average of the individual \vsys estimates obtained across all observing nights, derived by performing a least-squares fit of a Keplerian model to the multi-epoch radial velocity data. In this fit, all orbital parameters were fixed to literature values (see Table~\ref{tab_parameters}), and the \vsys was treated as the only free parameter, following the method proposed by \citet{Silva2024}. The resulting systemic velocities are -1.1161$\pm$0.0011~\kms and -1.2097$\pm$0.0016~\kms for HARPS-N and ESPRESSO, respectively. 
For GIANO-B, no RV values are available. However, we recalibrated our observations using the telluric spectrum (see Sect.~\ref{reduction}), and the instrument’s stability, as estimated from the telluric lines, is approximately 0.5~\kms (per individual order). We therefore estimate that GIANO-B and HARPS-N are in the same reference frame within an uncertainty of about 0.5~\kms, and we assumed for GIANO-B the \vsys derived from the HARPS-N data. \\
For the planet’s Keplerian orbital motion, we explored a range of planetary radial-velocity semi-amplitudes from $0$ to $300$~\kms, in steps of 1~\kms. 
Despite the application of a polynomial filter, order by order, to the $CC2D$ matrices, some stellar residuals remain. These non-stationary residuals, which can not be adequately modelled by a polynomial function, are likely due to the combined and time-variable contribution of both stars in the system, as the secondary star enters or exits the slit (or fibre) during the observations.
Since we are unable to fully remove these non-stationary residuals, we opted to mask the region of the cross-correlation function around the stellar position (see the masked band in the gray-scale $CC$ maps in Figs.~\ref{CO_Fe} and \ref{Fe_espresso}).
 We quantified the S/N of our CC map as in \citep[e.g.,][]{Brogi_2018_Giano, Giacobbe2021, Guilluy2022}. For each investigated molecule, we divided the total CC matrix by its standard deviation (calculated in the RV interval [-115, -35]~\kms and [+35, +115]~\kms to exclude the CC peak at the expected planet position).

\subsection{GCM calculation}
In this section, we model the phase-dependent thermal emission spectra of \planet\ to explore how our detection maps (in terms of \kp\ and \vrest) are influenced by orbital Doppler shifts and line strength variations. Previous pRT-based models, which assume static 1D atmospheres, have proven effective for first-order studies aimed at detecting atmospheric species. However, they overlook longitudinal temperature gradients and phase-dependent variations in spectral line profiles. By adopting more realistic, 3D-informed models, we aim to better interpret the observed emission and assess how atmospheric dynamics shape the detectability of species like CO and Fe~I.
We follow the methodology described in \citet{Wardenier2025} to generate phase-dependent emission templates for Fe~I and CO, based on a drag-free SPARC/MITgcm simulation of \planet (see also \citealt{Wardenier2021,Wardenier2023}). For details regarding the 3D temperature structure, chemistry, and dynamics, we refer the reader to \citet{Wardenier2025}; here, we summarize the key aspects.
The SPARC/MITgcm \citep{Showman2009} is a non-grey global circulation model that has been used extensively to simulate the 3D climates of (ultra-)hot Jupiters. The dayside TP profile -- shown in Figures 2 and 3 of \citet{Wardenier2025} -- exhibits a strong thermal inversion driven by the absorption of stellar irradiation by metals, with temperatures reaching up to $\sim$3500~K. This is consistent with the TP profile we used to compute 1D pRT model (Fig.~\ref{TP_profile}). Chemical abundances in the GCM are calculated assuming local chemical equilibrium and solar metallicity.
To obtain thermal emission spectra, we post-process the model with the 3D radiative transfer code gCMCRT \citep{Lee2022}. Templates are calculated at 10-degree phase intervals at a resolving power $R$ = 300,000 and then convolved with the spectrograph instrument profile (following the same approach used for the 1D models)\footnote{Since the grid of models is uniformly calculated at 10-degree phase intervals, for each observation in our dataset, we selected the grid model with the closest phase to that of the observation.}. Furthermore, we only include the opacities of the relevant line species, either Fe~I or CO, and the continuum (we used the same linelists as used for pRT, see Sect.~\ref{cc}).
For each species, we compute two sets of templates: one accounting for Doppler shifts due to planet rotation and the wind profile  (``3D templates with dynamics''), and one in which these shifts are disabled (``3D templates without dynamics''; see the first row in Fig.~4 in \citealt{Wardenier2025}). The latter set still accounts for changes in the 3D temperature structure during the observation, but not for the changes in Doppler shift introduced by the planet's atmosphere itself. The modelled wind field includes a day-to-night flow and a super-rotating equatorial jet, with wind speeds reaching up to $\sim$10~km/s.

\begin{figure*}
	\centering
        \includegraphics[width=\linewidth]{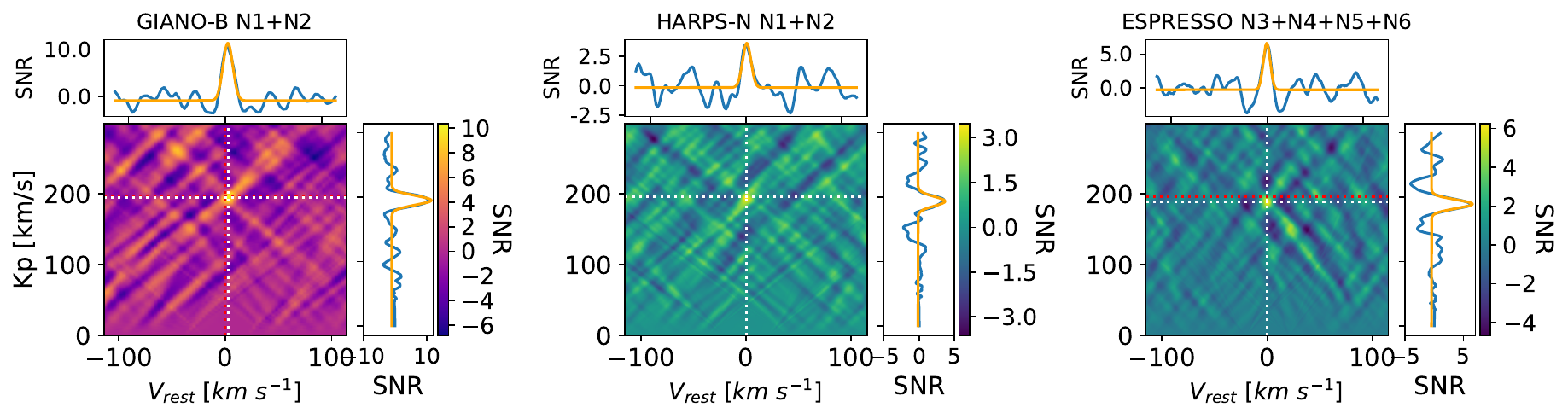}
	\caption{Detection S/N maps as a function of \vrest and \kp for CO in GIANO-B (left panel), Fe~I in HARPS-N (central panels), and Fe~I in ESPRESSO (right panel) data. Red (white) dotted lines mark the expected (obtained) planetary position. For each map, the top and right sub-panels show the 1D CC functions (in terms of S/N) at the peak position, with the best-fit Gaussian overplotted in orange}. 
	\label{detections}
\end{figure*}

\section{Results} \label{results}
In the present section, we summarise the results of our analysis. The main findings are reported in Table~\ref{tab_results} and illustrated in Fig.~\ref{detections}.
\begin{table*}[h!]
\centering
\caption{Cross-correlation results.}
\begin{tabular}{l|cc|cc|c}
\hline\hline
 & \multicolumn{2}{c|}{$V_{\text{rest}}$ [\kms]} & \multicolumn{2}{c|}{$K_p$ [\kms]} & S/N \\
 & gaussian fit & max(S/N)-1 & gaussian fit & max(S/N)-1 & \\
\hline
CO (GIANO-B)  & $2.71 \pm 0.51$ & $3.00_{-6.00}^{+3.00}$ & $195.00 \pm 0.34$ & $195.00_{-5.00}^{+5.00}$ & 10.4\\
Fe~I (GIANO-B)$^{*}$ &$14.5 \pm 0.5$  &     /       & $205.9 \pm 0.7$ & / & 4.0 \\
Fe~I (HARPS-N)  & $0.57 \pm 0.71$ & $0.00_{-4.50}^{+4.50}$ & $193.54 \pm 0.57$ & $196.00_{-9.00}^{+5.00}$ & 3.5 \\
Fe~I (ESPRESSO) & $-0.36 \pm 0.40$ & $0.00_{-3.00}^{+3.00}$ & $189.10 \pm 0.49$ & $189.00_{-5.00}^{+5.00}$ & 6.2 \\
\hline
\end{tabular}
\tablefoot{For each investigated molecule, the planet's orbital velocity ($K_p$), the velocity in the planet's rest frame ($V_{\text{rest}}$), and the S/N of the detection are reported. Two types of values (with associated uncertainties) are provided for each measurement: one obtained by fitting a Gaussian to the 1D cross-correlation function (at the peak position) and one estimated directly from the position of the peak of S/N in the map. In the latter case, the uncertainty was estimated by identifying where the S/N value dropped by 1 unit. As a consequence, the number of decimal digits used in the table differs depending on the method employed to derive the corresponding value.\\
   $^*$ Marginal detection obtained with only one observing night (namely N2).  The \kp and \vrest values have been computed by using a 2D Gaussian including a rotation angle accounting for the inclined feature of the detection blob (see Sect.~\ref{gianob_res}).}
\label{tab_results}
\end{table*}

\subsection{Near-infrared: GIANO-B} \label{gianob_res}
We report a detection of CO in the GIANO-B data (see the first panel of Fig.~\ref{detections}) with a S/N of 10.4. The planetary signal is clearly visible in both nights, with S/N map peaks of 5.0 and 7.2 for Night~1 and Night~2, respectively (see Fig.~\ref{CO_Fe}). This result confirms the detection previously reported by \citet{Yan2023} using CRIRES+, which was based on single-night observations.\\
We also searched Fe~I in our GIANO-B data; however, we were not able to detect a clear peak in the cross-correlation map (see right panels of Fig.~\ref{CO_Fe} in the appendix). There seems to be only a hint of signal during the second night of observation at the expected position for a signal of planetary origin. The peak of the S/N map is located at 
\vrest = $14.5 \pm 0.5$ \kms and \kp =$205.9 \pm 0.7$ \kms, obtained by fitting a 2D Gaussian that includes a rotation angle to account for the tilted shape of the signal trace and reaches a maximum S/N of 4.0. Since we are probing less than a quarter of the planet’s orbit, we have the characteristic “degeneracy” between \kp and \vrest, although the theoretical value still lies along the trail of the peak. 
Despite a S/N of 4.0, we refer to this detection as "marginal" because it is present only in one of the two GIANO-B nights. Here, "marginal" does not imply that the signal is statistically weak, but rather emphasises two factors: the detection occurs in only one epoch, and the peak in the S/N map is slightly offset from the expected theoretical position.

\subsection{Optical: HARPS-N and ESPRESSO}
Regarding the analysis in the visible band, we detected Fe~I in both HARPS-N (see the central panel of Fig.~\ref{detections}) and ESPRESSO (see the right panel of Fig.~\ref{detections}) spectra with an S/N of 3.5 and 6.2, respectively.

\citet{Silva2024} first analysed ESPRESSO data of \planet's dayside and reported a consistent blue shift of -4.7 $\pm$ 0.3~\kms\ in the Fe~I line. However, our re-analysis does not confirm this result, as our measured V$_\mathrm{rest}$ is consistent with 0~\kms\ (see Table ~\ref{tab_results}). Our findings appear to be more in line with the GCM predictions by \citet{Wardenier2025}, which suggest a negligible $\Delta$\vrest when combining pre- and post-eclipse phases. Both the right panel of Fig.~\ref{detections} and Table~\ref{results} show that the measured \kp exhibits a negative offset relative to the theoretical value of 196.52$\pm$0.94~\kms. 

In the analysis of the ESPRESSO data, we adopted a different strategy compared to \citet{Silva2024}, particularly in the pre-CC processing. Rather than applying the {\tt Molecfit} tool \citep{2015A&A...576A..77S, 2015A&A...576A..78K} for telluric correction, as done by \citet{Silva2024}, we employed the PCA, a technique widely adopted in the literature for cross-correlation-based studies. Moreover, while \citet{Silva2024} used the RASSINE code \citep{Cretignier2020} to remove the continuum and instrumental interference patterns (commonly referred to as “wiggles”), we applied a high-pass filter to eliminate low-frequency systematics that are not stable over time. We adopted a window of approximately 70~\kms, chosen to be wide enough not to affect the planetary signal, yet narrow enough to correct for the wiggle frequencies \citep{Bourrier2024}. We verified through Lomb-Scargle periodograms that no significant periodicities remained in the post-PCA residuals. Our pre-CC methodology follows the standard framework originally developed for nIR data, which we here adapt and implement for optical spectroscopy. Furthermore, \citet{Silva2024} employed an unweighted cross-correlation approach, whereas we followed the formalism of \citet{Gibson2020}, \citet{Cont2022}, and \citet{Nortmann2025} to account for the uncertainties on the post-PCA residual spectra. To validate the performance of our analysis pipeline, we compared the cross-correlation maps, S/N maps, and final detections -- using both injection tests and real data -- with the results obtained from an independent tool previously adopted by our team for HARPS-N data analysis (see, e.g., \citealt{Borsa2022}). The two methods yielded fully consistent results.

\subsection{GCMs results}
\citet{Wardenier2025} demonstrated how 3D atmospheric modelling can account for small offsets in the inferred orbital velocity $K_p$ observed in emission spectroscopy, attributing these discrepancies to the three-dimensional structure of WASP-76b’s atmosphere. Motivated by their findings, we decided to test the application of GCMs in our analysis as well.
We decided to test the GCMs on our most robust detections, namely CO in the GIANO-B spectra and Fe~I in the ESPRESSO data.
As highlighted in Sect.~\ref{data_analysis}, we investigated two classes of atmospheric models: one excluding dynamical effects but including only a phase-dependent planetary atmosphere, i.e., a rotating 3D temperature structure without Doppler shifts, and one incorporating both wind and rotational broadening (see Fig.~\ref{GCM_comp}). Within the dynamic framework, we compared a weak drag model (with a drag timescale of 10$^5$ seconds; not shown here for conciseness) and a drag-free model (which turned out to show the best agreement with the data). We adopted a likelihood-based comparison approach to evaluate which model is most consistent with the data.
Specifically, we transformed the CC values into likelihood (LH) values following the formalism introduced by \citet{Gibson2020} and later adopted by \citet{Giacobbe2021}. In this approach, the likelihood incorporates both the model line depths and the S/N across spectral orders and exposures. Detecting a peak in the LH map at the expected (\vrest, \kp) coordinates serves as a robust confirmation of the planetary signal. For each model, we computed a log-likelihood map over the (\vrest, \kp) parameter space by summing the contributions from each spectral order, observation, and RV shift, scanning  \vrest values from $-5$ to $+5$~\kms in steps of 0.3~\kms, and \kp from 186~\kms to 202~\kms in steps of 0.3~\kms. To ensure consistency with our data reduction, the same telluric correction and PCA-based filtering applied to the observations were also applied to the model spectra. Finally, confidence levels on the detection were derived using the likelihood-ratio test, comparing each point in the (\vrest, \kp) grid to the maximum LH value. 

\begin{figure}
    \includegraphics[width=1.05\linewidth]{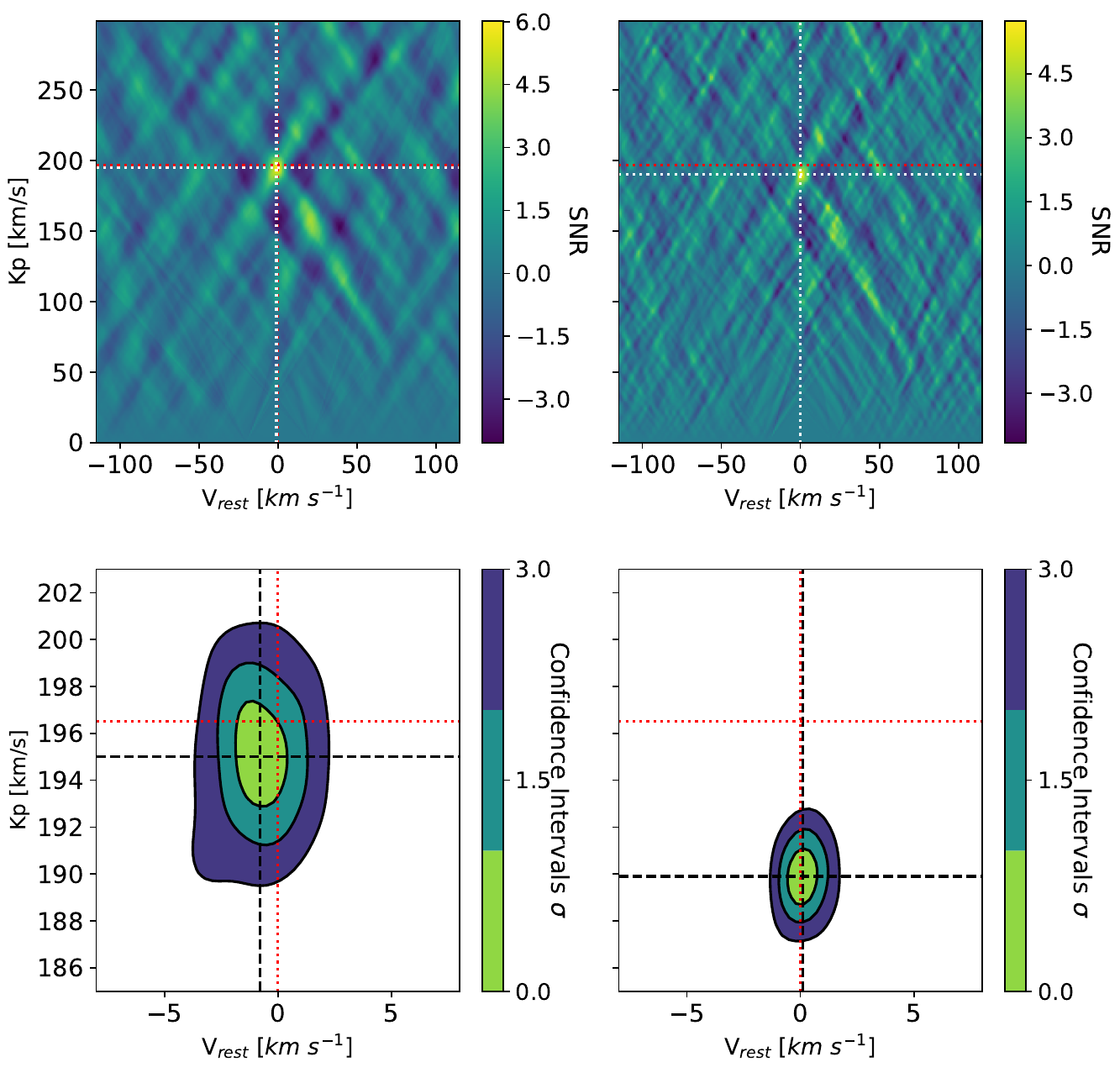}
    \caption{3D templates with dynamics (left panels) vs 3D templates without dynamics (right panels) comparison for Fe~I in ESPRESSO data. Upper panels: detection S/N maps as a function of \kp and \vrest with the 1D CC functions (see Fig~\ref{detections} for details). Bottom panels: Likelihood confidence intervals as a function of \kp and \vrest. Red (black) dotted lines mark the expected (obtained) planetary position. The 3D templates were computed based on the drag-free GCM of \planet from \citet{Wardenier2025}. }
    \label{GCM_espresso}
\end{figure}

 \noindent{The} results are shown in Fig.~\ref{GCM_espresso}. Although there is no significant advantage in terms of LH between the dynamic and non-dynamic 3D models (the non-drag dynamic model is only favoured by $\delta\log(L)=2$ compared to the drag-free model without Doppler effects), applying 3D GCMs does yield a clear improvement in \kp-\vrest space. In the case of no-dynamic models, the LH peak is incompatible with the theoretical planetary position. Even when considering the uncertainty of about 1~\kms on the theoretical \kp (see Table~\ref{tab_parameters}), the 3$\sigma$ interval does not overlap with the 3$\sigma$ confidence range of the model without dynamics.\\ Conversely, when adopting GCMs with atmospheric dynamics, i.e. the drag-free model, the LH peak becomes consistent with the expected \vrest, \kp values within 1 to 2$\sigma$ (if we account for the error of 1~\kms on the theoretical \kp). The physical aspect of the GCM responsible for the better fit is the fact that we account for phase-dependent Doppler shifts due to planet rotation and the wind profile (see Fig.~11 in \citealt{Wardenier2025}). For completeness, we also tested the weak-drag dynamical model, which resulted in a detection compatible with the theoretical \kp and \vrest values at the 3$\sigma$ level. However, because the agreement between this model and our data is worse compared to the drag-free model, we do not further discuss it here. Our result is in agreement with \citet{ehrenreich2020}, who pointed out that an offset of the hotspot is in tension with the possible existence of strong drag forces.
 
\noindent{Regarding} the application of GCMs to our GIANO-B data, our results are inconclusive, as both the drag-free and the without-dynamics models are consistent within 1$\sigma$ with the theoretically expected position for a planetary signal, possibly due to the limitations imposed by the lower S/N \citep{Pluriel2023}. 
However, it is worth noting that, according to Table~\ref{tab_results}, the observed \kp for CO is consistent with the expected \kp from the literature within the uncertainties. Consequently, there is no conclusive evidence for a non-zero $\Delta$\kp, and thus for significant 3D effects. To better assess the role of dynamics, more precise measurements of \kp from CO cross-correlation analysis would be required.

\begin{table*}[]
\centering
\caption{GCM results}
\begin{tabular}{l|cc cc c|cc}
\hline\hline
& \multicolumn{5}{c|}{Cross-correlation framework} & \multicolumn{2}{c}{Likelihood framework}  \\
\hline
 & \multicolumn{2}{c}{\vrest [km\,s$^{-1}$]} 
 & \multicolumn{2}{c}{\kp [km\,s$^{-1}$]} 
 & S/N 
 & \vrest [km\,s$^{-1}$] & \kp [km\,s$^{-1}$] \\
 & Gaussian fit & max(S/N)-1 
 & Gaussian fit & max(S/N)-1 
 &  
 &  &  \\
\hline
3D templates with dynamics
& $-1.38 \pm 0.41$ & $-1.00_{-2.00}^{+3.00}$ & $194.57 \pm 0.49$ & $195.00_{-5.00}^{+5.00}$ & 6.0
& $-0.80_{-1.20}^{+1.50}$ & $195.00_{-2.40}^{+2.40}$\\
3D templates without dynamics
& $0.57 \pm 0.34$ & $0.00_{-2.00}^{+2.00}$ & $189.43\pm 0.49$ & $190.00_{-4.00}^{+3.00}$ & 5.7 
& $0.10_{-0.90}^{+0.90}$ & $189.90_{-1.50}^{+1.20}$ \\
\hline
\end{tabular}
    \tablefoot{For each tested model, the velocity in the planet's rest frame (\vrest), the planet's orbital velocity ($K_p$) and the S/N of the detection are reported. Two types of values (with associated uncertainties) are provided for each cross-correlation measurement: one obtained by fitting a Gaussian to the 1D cross-correlation function, and one estimated from the S/N as the peak value minus one. For the likelihood result, we reported the \kp and \vrest of the peak position.}
    \label{GCM_e}
\end{table*}

\section{Discussion and Conclusion} \label{conclusions}

In this paper, we analysed two observations of the dayside of \planet obtained with the GIARPS observing mode of the TNG, which combines the two high-resolution spectrographs GIANO-B in the nIR and HARPS-N in the optical. Additionally, we reanalysed four nights of publicly available ESPRESSO data, which were initially examined in \citet{Silva2024}.  

Our primary goal was to study multi-band spectra to achieve the best possible characterisation of CO and Fe~I, two key tracers of the dynamics and thermal structure of the planetary dayside. A crucial aspect of our work was applying a consistent analysis approach to both nIR and optical datasets.  

We detected CO in our GIANO-B spectra with an S/N of 10.4, confirming the previous detection obtained with CRIRES+ by \citet{Yan2023}. Additionally, we report the detection of Fe~I with S/N values of 3.5 and 6.2 for  HARPS-N and ESPRESSO, respectively. However, for ESPRESSO data, we were unable to reproduce the blue-shifted signal reported by \citet{Silva2024}. The retrieved values of \kp and \vrest are consistent with theoretical predictions or show a small deviation in $\Delta$\kp ($\Delta$\kp$\sim$6~\kms compatible with the theoretical \kp of $\sim196$ \kms, see Table~\ref{tab_parameters}, at the $\sim$4$\sigma$ level - value obtained by averaging the \kp values and their respective uncertainties listed in Table~\ref{detections}), in agreement with \citet{Wardenier2025}.  \\
A "marginal" detection of Fe~I is also suggested in the GIANO-B data (see right panel of Fig.~\ref{CO_Fe} in the appendix), although the signal is embedded in a noisy map. However, we verified through injection-recovery tests that the spurious peaks are unlikely to be caused by cross-contamination from other molecular or atomic species (i.e., Cr, H$_2$O, OH, CO, CO$_2$, Mg). This is an important check, especially in the case of a few Fe~I lines. While this may not yet qualify as a robust detection, we observe that the LH significance of the Fe~I detection improves when combining the most favourable GIANO-B night with the HARPS-N data (see Fig.~\ref{Fe_giano}). Although the detection is still primarily driven by HARPS-N, GIANO-B contributes by cleaning up the signal and reducing the presence of spurious peaks in the LH map. We stress that, even though we consider the Fe~I detection with GIANO-B as "marginal" for the reasons explained in Sect.~\ref{gianob_res}, its genuineness is somewhat strengthened by the same detection with HARPS-N. Specifically, GIANO-B confirms the same features that HARPS-N detected.

\noindent{Furthermore}, in this work, we tested GCM-based models for CC, particularly to verify the theoretical predictions that GCMs can resolve the small discrepancies in the retrieved \kp values typically observed when using simplified 1D models. This study represents one of the few applications of GCMs both in CC and LH frameworks on real data (see, e.g., \citealt{Flowers2019}, \citealt{Beltz2021}). Although, from a statistical perspective, the LH framework favours the dynamic GCM by only $\delta\log(L)=2$ in our ESPRESSO analysis (see Table~\ref{GCM_e}), the LH maps in Fig.~\ref{GCM_espresso} clearly show that—at the $2\sigma$ level—only the dynamic model aligns with the expected planetary velocity. In contrast, a static model fails to reproduce this offset. We find that only the drag-free GCM is able to produce the magnitude of the \kp offset seen in the ESPRESSO data. A weak-drag model (with lower wind speeds and no super-rotating equatorial jet) cannot fully reproduce the size of the \kp offset seen in the data.

We noted that in both HARPS-N and GIANO-B data, the planetary iron signal appears to be stronger during post-eclipse, with S/N and likelihood maps showing fewer spurious features during the post-eclipse night (see right panels of Fig.~\ref{CO_Fe} for GIANO-B and Fig.~\ref{like_harps} for HARPS-N in the appendix). 
Although the CO signal is detected in both nights, some effects --possibly instrumental issues or weather conditions-- may have affected our measurements during the first GIARPS observation. However, since CO remains detectable in Night~1 and the night logs report no significant differences in weather conditions, the origin of this nightly variation remains unclear. Given the intrinsically weaker detection of the Fe~I signal compared to CO, it is still plausible that some instrumental issues may have impacted our GIANO-B results in Night~1.
It is worth noting that \citet{Yan2020} observed a similar effect in the atmosphere of WASP-189\,b, with a stronger Fe~I signal just after the secondary eclipse event. However, they attributed this difference to a lower S/N in the pre-eclipse observation, likely caused by a telescope focusing issue. According to Fig.~\ref{observations},  also in our GIARPS observations, the strongest detection corresponds to a higher S/N ratio. 
Interestingly, our re-analysis of the ESPRESSO data reveals that also in these observations, Fe~I appears weaker in pre-eclipse than in post-eclipse observations, see Fig.~\ref{pre_post_maps} and Table~\ref{pre_post_table}. 

\noindent{The} stronger Fe~I detection at post-eclipse phases is consistent with the idea that the hotspot is shifted from the substellar point toward the evening terminator, as often seen in phase curve observations \citep[e.g.,][]{wong2016}. 
 As shown in \citet{vanSluijs2023} and \citet{Wardenier2025}, the vertical temperature gradient is stronger outside the hotspot than inside. Because the strength of the emission lines is set by the temperature gradient (and not by the absolute temperature), we expect the emission signal to be stronger than in pre-eclipse (in which most of the hotspot is in view). A similar explanation was also given for Kelt-20\,b by \citet{Borsa2022}. The authors found an asymmetric detection for Fe~II and Cr~I detected only after the occultation and not before, hinting at different atmospheric properties in view during the pre- and post-occultation orbital phases.
 
\noindent{This} asymmetry in \planet's atmosphere, may indicate the presence of zonal flows that transport heat longitudinally in a prograde direction (in the same direction as the tidally locked solid-body rotation, $\lambda = 61.28^{+7.61}_{-5.06}$~deg; \citealt{ehrenreich2020}) in the upper atmosphere, which would shift the hotspot position eastward. We may observe this effect only with Fe~I (and not with CO) because Fe~I forms higher up in the atmosphere at higher temperatures. 
This is largely consistent with the findings of \citet{ehrenreich2020}, who introduced a hotspot offset toward the evening terminator to explain the asymmetry observed between the ingress and egress phases of \planet's transit. It is also consistent with theoretical studies (e.g., \citealt{ShowmanPolvani2011} and \citealt{Parmentier2018_book}), which suggested that eastward shifts in thermal phase curves are a robust outcome of hot Jupiter circulation regimes. However, \citet{May2021} analysed the IR phase curves of \planet and found little to no asymmetry, a result confirmed by a reanalysis of \citet{Demangeon2024} and of \citet{Dang2025}. Notably, \citet{Demangeon2024} also examined phase curves in the optical regime using TESS and CHEOPS data and identified a peak flux excess before the eclipse, suggesting a tentative asymmetry in the phase curve at visible wavelengths. This asymmetry could indicate an asymmetric wind pattern from west to east in the atmospheric layers probed at visible wavelengths. As also noted by \citet{May2021}, the authors suggested that the dynamics of atmospheric layers probed at visible and IR wavelengths may differ. An alternative explanation for the hint of asymmetry in the optical light curves invoked by \citet{Demangeon2024} is scattering, the "glory effect" and associated clouds in the eastern hemisphere. However, this hypothesis is challenged by theoretical predictions that clouds should form in the western hemisphere (e.g., \citealt{Hu2015}; \citealt{Parmentier2016}; \citealt{Lee2016}; \citealt{Helling2019}). In Fig.~\ref{hotspot}, we graphically illustrate the two different explanations for the asymmetry in the Fe~I detection between the pre- and post-eclipse phases that we highlighted in this section.
From a preliminary analysis of JWST/NIRSpec data (private communication, JWST GO 5268, PI: Wardenier), non-zero eastward offsets of $\sim$5~degrees appear to be present, reinforcing our hypothesis that a difference in the temperature gradient may cause the Fe~I asymmetry we hint at between the pre- and post-eclipse phases.

This work represents one of the first steps toward conducting similar studies on a larger sample of exoplanets within the framework of the GAPS program.
The GAPS program at the TNG has collected multiple transits (ranging from 3 to 11) for a key sample of approximately 40 exoplanets using GIARPS, building a unique and unparalleled dataset for the astronomical community in the coming years. This dataset will enable detailed studies on the repeatability of atmospheric detections and the variability of exoplanetary atmospheres, laying the groundwork for future investigations with the ELT. 
 Furthermore, we plan to combine our observations with publicly available archival ESPRESSO and CRIRES+ data to take advantage of the highest-resolution spectrograph currently available. 
  In cases of high S/N detections, we will also be able to test GCMs when available, as demonstrated in this work. Observational constraints are essential for GCMs. For instance, to measure the impact of drag would require more observations of HJ and UHJs to be fully modelled. The detection of species with accurate abundances would also help to determine the atmospheric regime (e.g., equilibrium, non-equilibrium, etc.), which has a significant impact on GCMs, as shown in \citet{Pluriel2023}.

\bibliographystyle{aa}
\begin{acknowledgements}
We thank the anonymous referee for their valuable feedback on this paper. We are also grateful to A.~Chiavassa for his kind and insightful support in our efforts to model the stellar contamination. We acknowledge the Italian Centre for Astronomical Archives (IA2, https://www.ia2.inaf.it), part of the Italian National Institute for Astrophysics (INAF), for providing technical assistance, services and supporting activities of the GAPS collaboration. We acknowledge A.~Lanza for very helpful comments on the manuscript. J.P.W. acknowledges support from the Trottier Family Foundation via the Trottier Postdoctoral Fellowship held at IREx and the Canadian Space Agency (CSA) under grant 24JWGO3A-03. The authors acknowledge financial contributions from PRIN INAF 2019 and INAF GO Large Grant 2023 GAPS-2 as well as from the European Union - Next Generation EU RRF M4C2 1.1 PRIN MUR 2022 project 2022CERJ49 (ESPLORA). P.E.C. is funded by the Austrian Science Fund (FWF) Erwin Schroedinger Fellowship, program J4595-N. Part of the research activities described in this paper were carried out with contribution of the Next Generation EU funds within the National Recovery and Resilience Plan (PNRR), Mission 4 - Education and Research, Component 2 - From Research to Business (M4C2), Investment Line 3.1 - Strengthening and creation of Research Infrastructures, Project IR0000034 – “STILES - Strengthening the Italian Leadership in ELT and SKA”.
\end{acknowledgements}
\bibliography{ref_w76}

\begin{appendix} \label{appendix}

\clearpage
\onecolumn
\section{Additional figures and tables} \label{add}

\begin{table}[!htp]
\centering
\caption{Number of PCA components used in this analysis}
\label{pca_comp}
\begin{tabular}{l|c|c}
\hline\hline
Instrument & Night & \# PCA Components \\
\hline
\multirow{2}{*}{GIANO-B}   & N1 & 5 \\
                           & N2 & 6 \\
\hline
\multirow{2}{*}{HARPS-N}   & N1 & 3 \\
                           & N2 & 4 \\
\hline
\multirow{4}{*}{ESPRESSO}  & N3 & 3 \\
                           & N4 & 3 \\
                           & N5 & 3 \\
                           & N6 & 4 \\
\hline
\end{tabular}
    \tablefoot{From left to right, we report: the instrument; the night and the number of components used in the PCA analysis.}
\end{table}

\begin{figure}[!htp]
	\centering
    \subfigure[CO (GIANO-B)]{
        \includegraphics[width=0.45\linewidth]{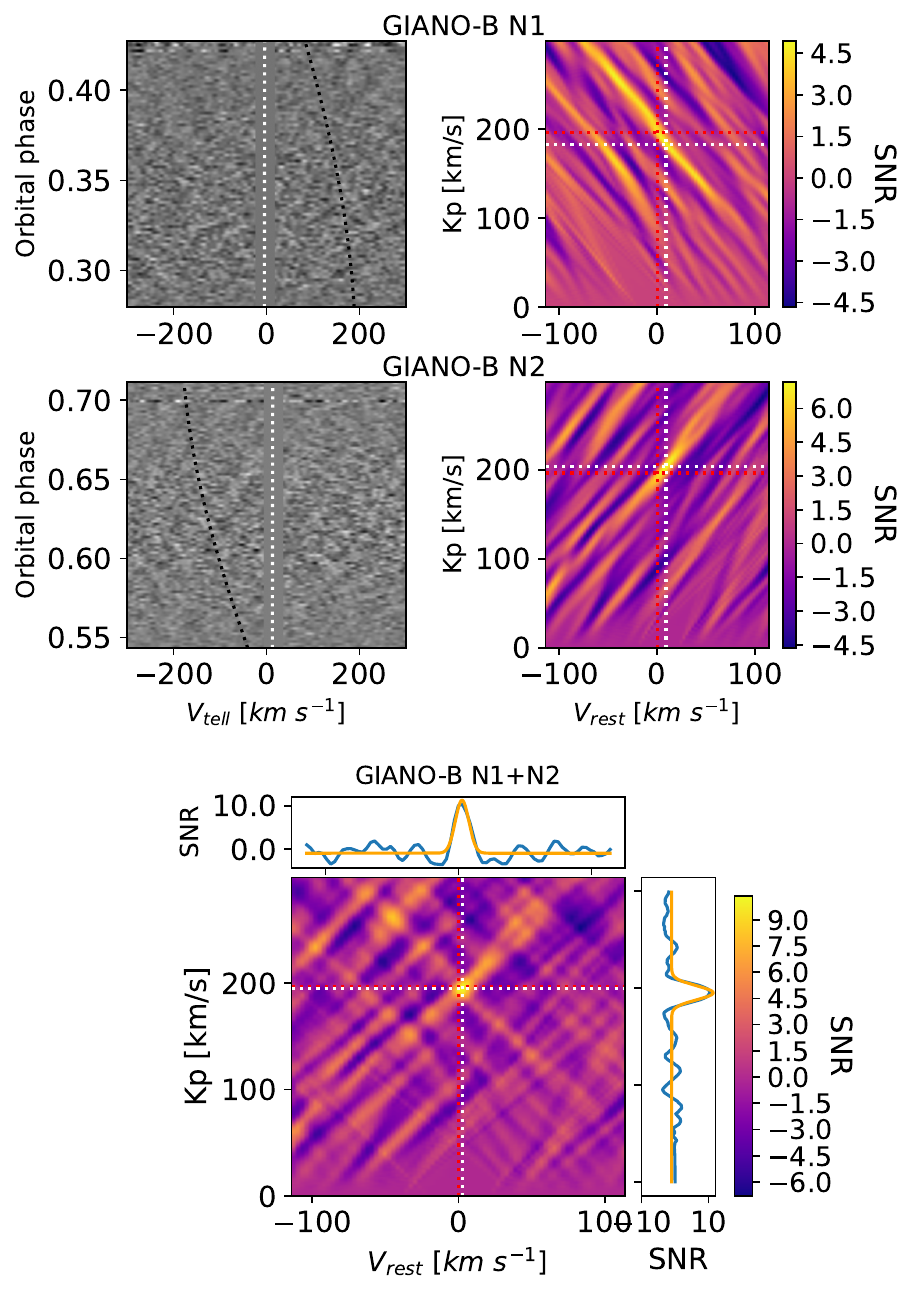}}
    \subfigure[Fe~I (GIANO-B)]{
        \includegraphics[width=0.45\linewidth]{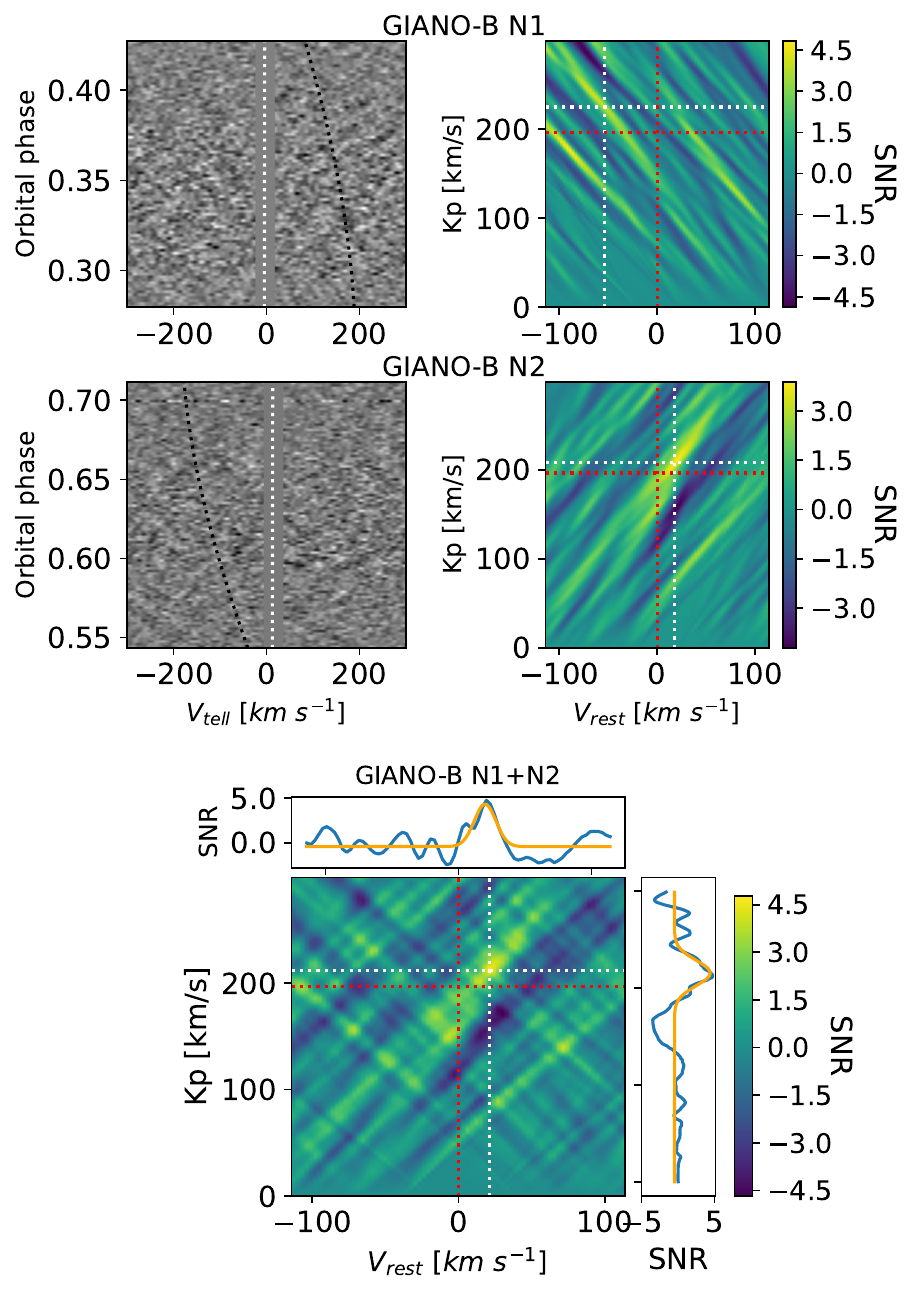}}
	\caption{Cross-correlation results for CO (left panels) and Fe~I (right panels) in GIANO-B data. For each species, we show the results for individual nights (top two rows) and their combined signal (bottom row). The greyscale maps represent the cross-correlation functions as a function of velocity in the telluric rest frame (\vtell) and orbital phase. Masked regions are affected by stellar residuals (see Sect.~\ref{data_analysis}). The black and white dashed lines indicate the stellar and planetary trails, respectively.\\ The colour maps display the 2D cross-correlation function in the (\kp, \vrest) plane, expressed in terms of S/N. A detailed description of these plots is provided in the caption of Fig.~\ref{detections}. Since we adopted the CC approach from \citet{Gibson2020}, \citet{Cont2022}, and \citet{Nortmann2025}, the noisier lines have larger uncertainties, and a bigger relative scatter among the data.}
	\label{CO_Fe}
\end{figure}

\clearpage

\begin{figure}
    \centering
    \subfigure[Fe~I (HARPS-N)]{
        \raisebox{5.4cm}{\includegraphics[width=0.45\linewidth]{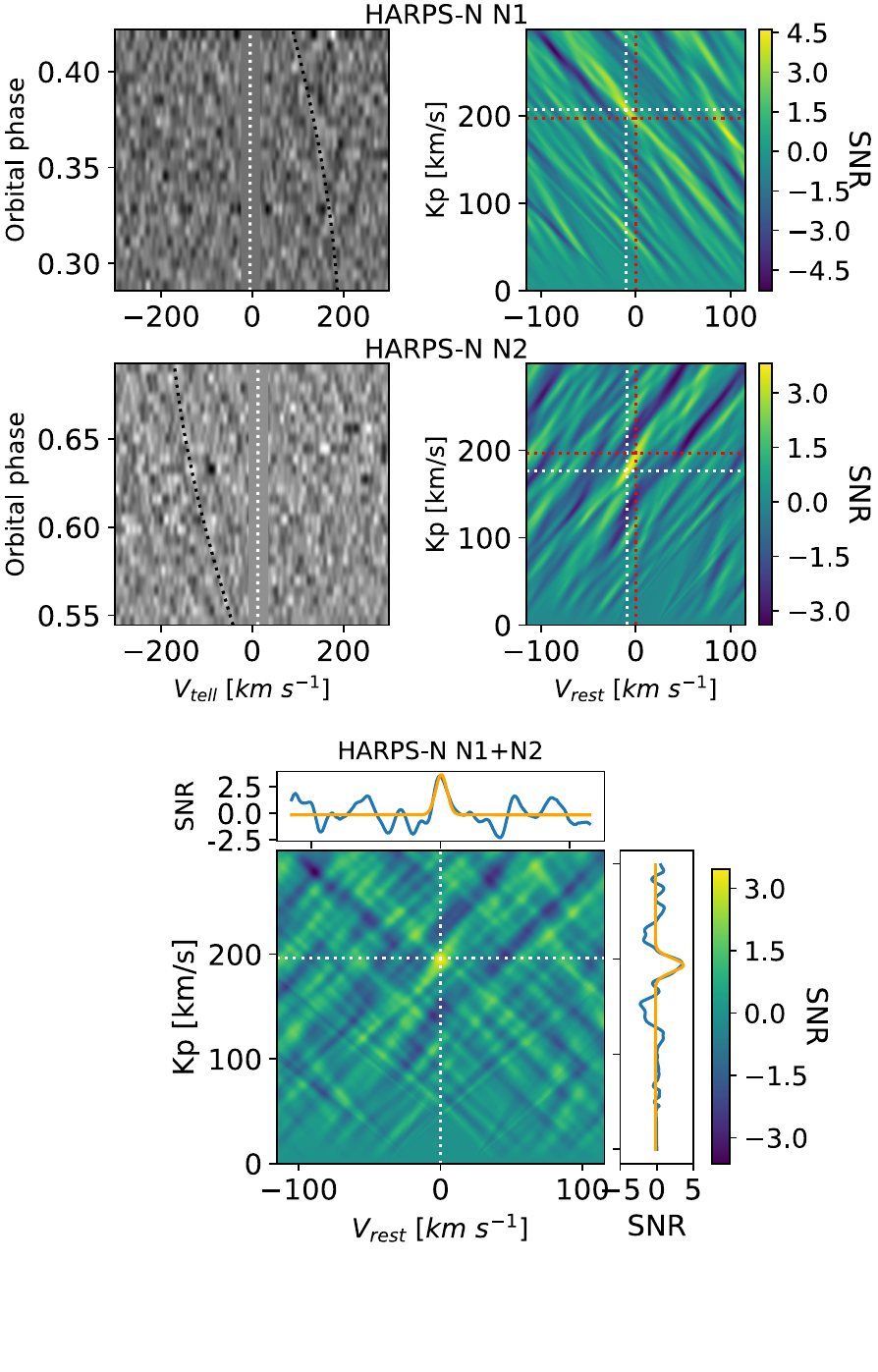}}}
    \subfigure[Fe~I (ESPRESSO)]{
        \includegraphics[width=0.5\linewidth]{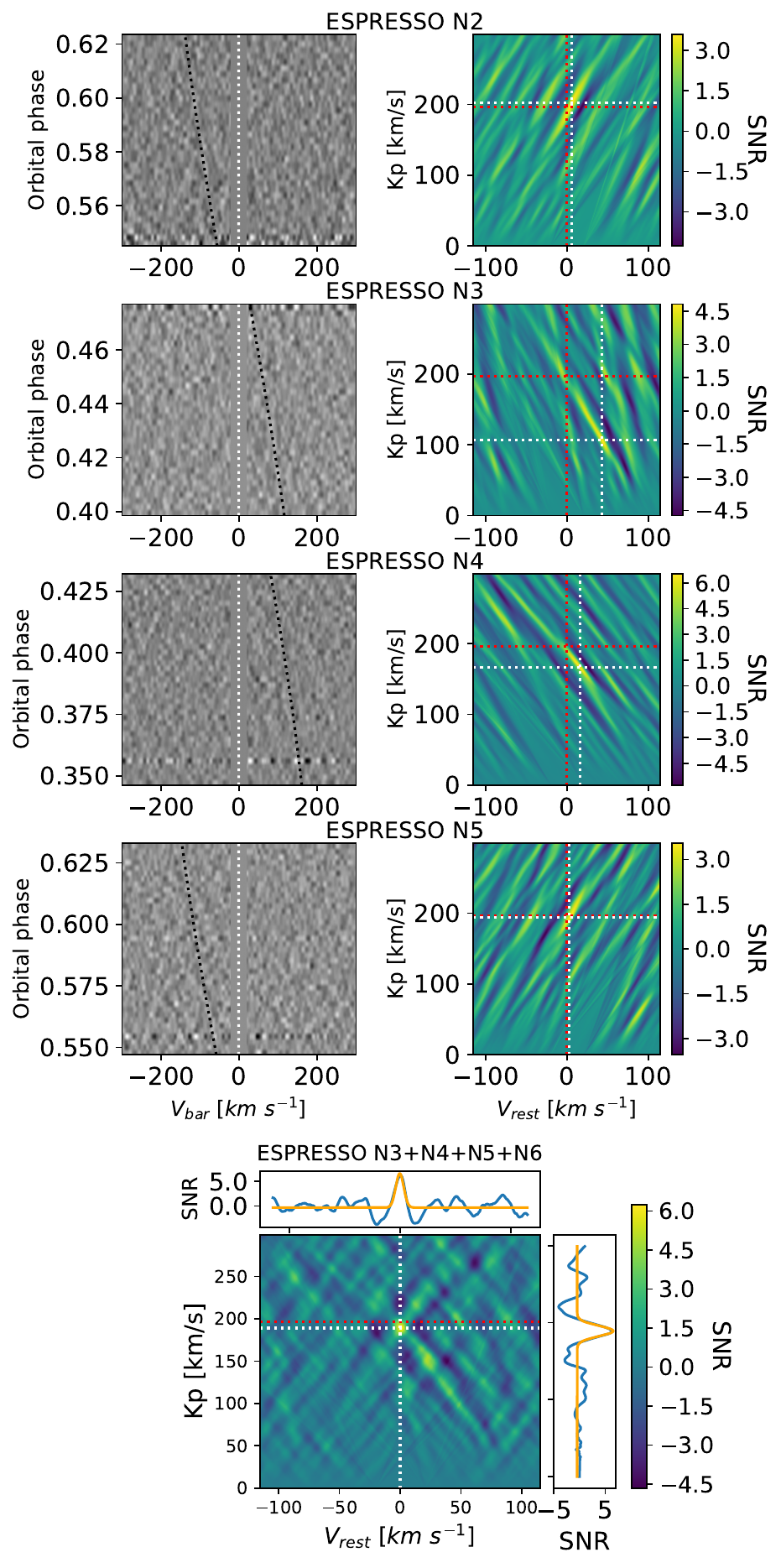}}
    \caption{Cross-correlation results for Fe~I in HARPS-N (left panels) and ESPRESSO (right panels) data. Same as Fig.~\ref{CO_Fe}. The ESPRESSO CC grey maps are in the barycentric rest frame.}
    \label{Fe_espresso}
\end{figure}

\twocolumn
\begin{figure}
	\centering
        \includegraphics[width=\linewidth]{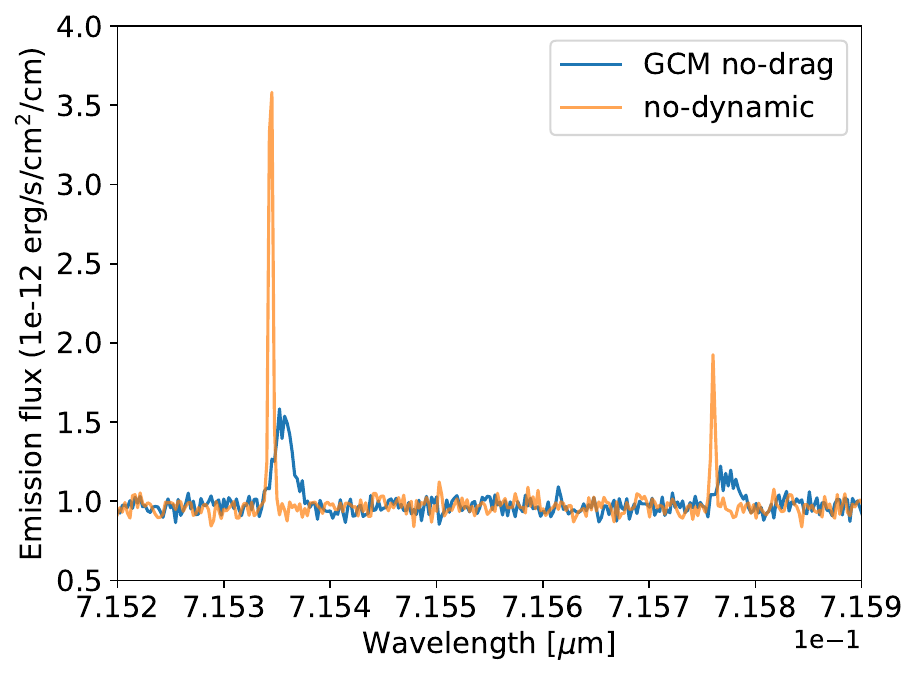}
	\caption{Emission flux as a function of wavelength, computed at a phase angle of approximately 0.583, for both the drag-free GCM (in blue) and the 'static' model (in orange), in a region around a Fe~I line.}
	\label{GCM_comp}
\end{figure}
\begin{figure}
	\centering
        \includegraphics[width=\linewidth]{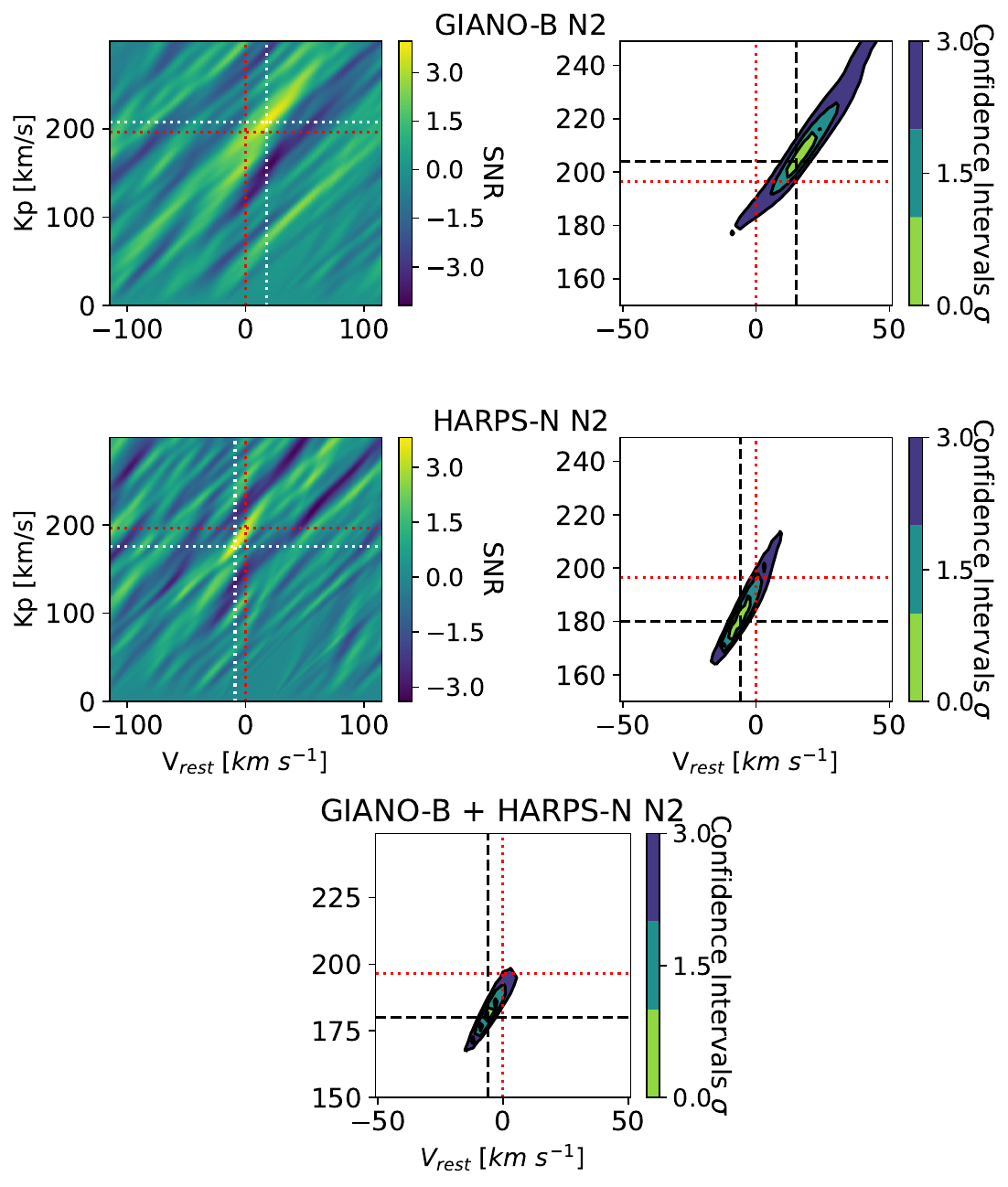}
        \caption{Fe~I detection with GIARPS during Night 2. For each instrument (top row: GIANO-B; middle row: HARPS-N), we present: the 2D CC maps in the (\kp, \vrest) plane, expressed in terms of S/N (left panels, colour scale); and the corresponding likelihood confidence interval maps (right panels). The bottom row shows the combined GIANO-B + HARPS-N confidence interval maps. Although the detection is primarily driven by HARPS-N, GIANO-B contributes by cleaning the signal and reducing spurious peaks in the likelihood maps.}
        \label{Fe_giano}
\end{figure}
\begin{figure}
	\centering
        \includegraphics[width=\linewidth]{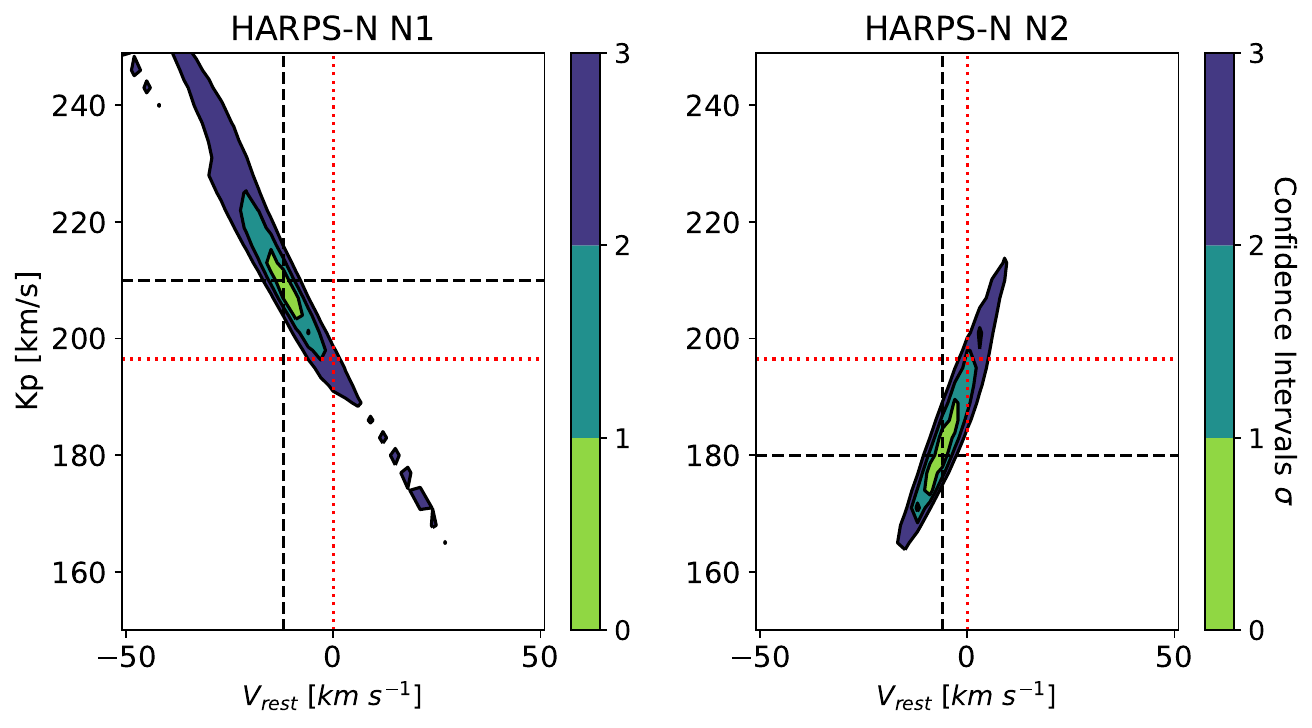}
	\caption{Likelihood confidence interval maps of Fe~I detection with HARPS-N as a function of \vrest and \kp. Red (black) dotted lines mark the expected (obtained) planetary position. }
	\label{like_harps}
\end{figure}

\begin{figure}[h]
\centering
\includegraphics[height=11cm]{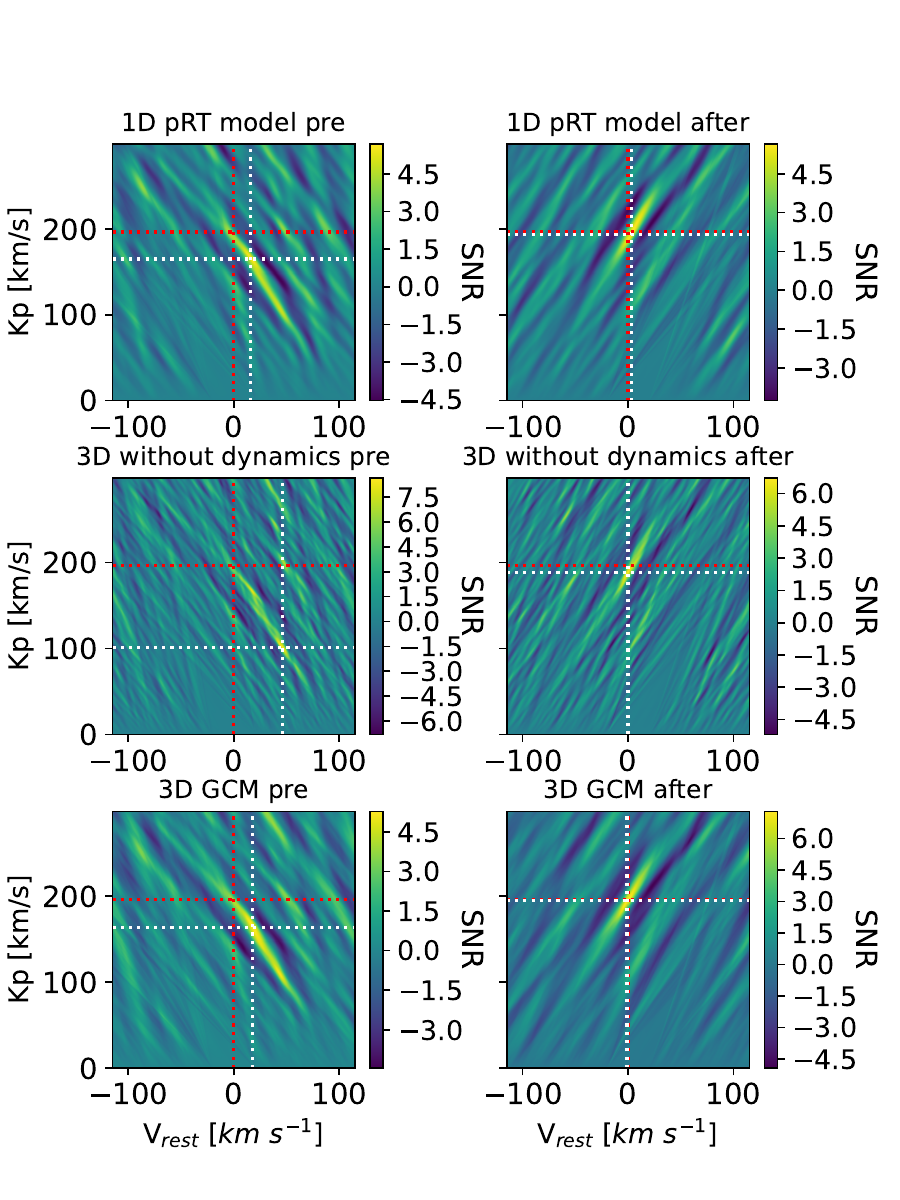}
\caption{Detection of Fe~I with ESPRESSO during pre-eclipse (left panels) and post-eclipse (right panels). For each atmospheric model (top row: 1D pRT; middle row: 3D without dynamics; bottom row: 3D no-drag GCM), we present the two-dimensional cross-correlation maps in the (\kp, \vrest) plane, expressed in terms of S/N (see Fig.~\ref{CO_Fe} for more details).}
\label{pre_post_maps}
\end{figure}

\begin{table}
\centering
\caption{S/N values measured at the expected planetary position (\kp$=196.52$~km~s$^{-1}$ and \vrest$=0$~km~s$^{-1}$) for each model in Fig.~\ref{pre_post_maps}.}
\label{pre_post_table}
\begin{tabular}{c|cc}
\hline \hline
Model & SN pre-eclipse & SN post-eclipse \\
\hline
1D pRT model        & 2.0 & 4.0 \\
3D without dynamics & 2.4 & 3.2 \\ 
3D GCM              & 3.6 & 7.3 \\ 
\hline
\end{tabular}
    \tablefoot{In all cases, the post-eclipse phase exhibits a higher S/N compared to the pre-eclipse phase.}
\end{table}

\begin{figure*}
	\centering
        \includegraphics[width=\linewidth]{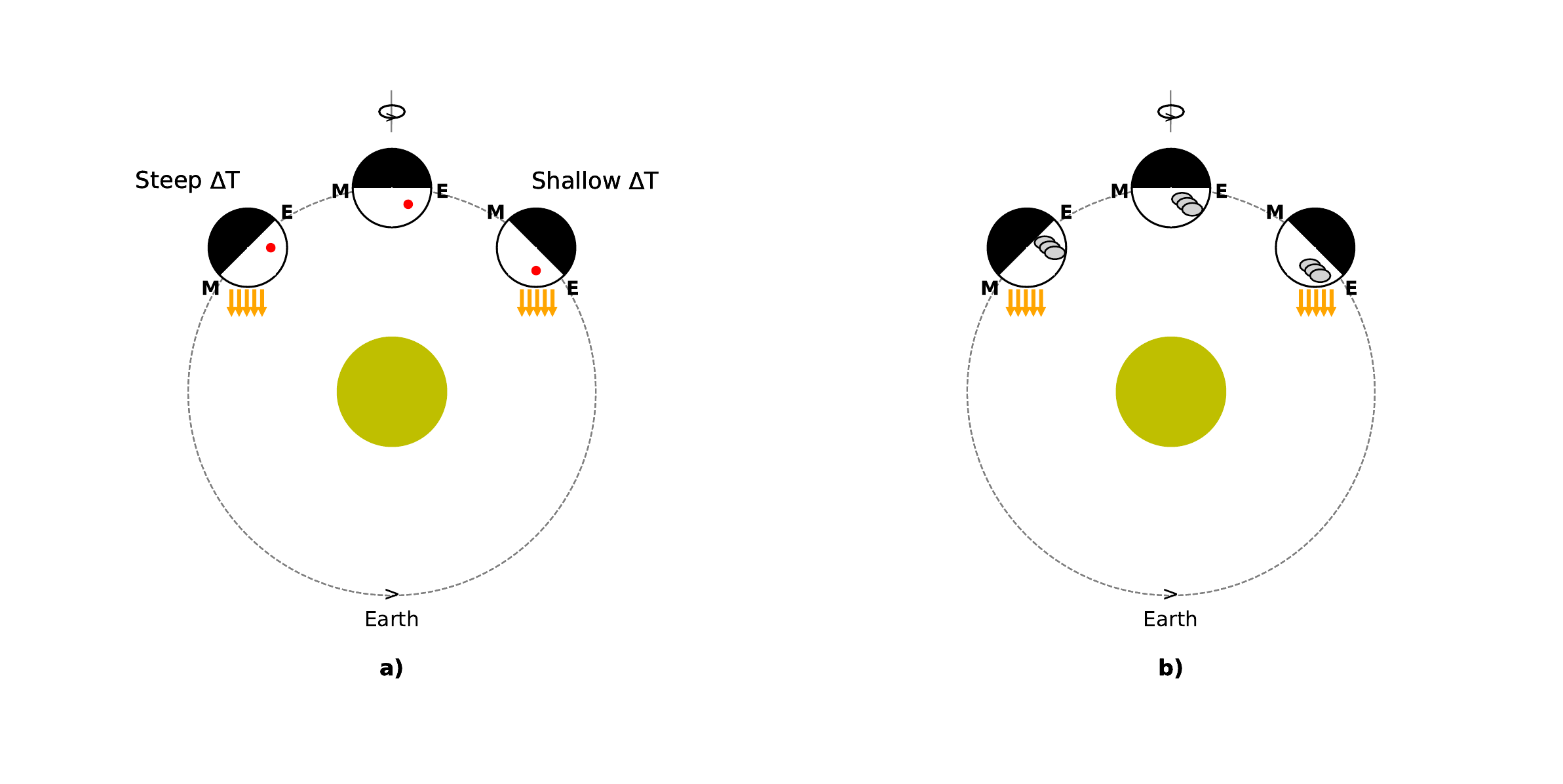}
	\caption{Possible explanation for the asymmetry of our detection. a) panel, the asymmetry could be due to the eastward shift of the hotspot. During the pre-eclipse phase (when we observe the eastern side with the hotspot), the observed temperature gradient is shallower.
The temperature gradient is steeper during the post-eclipse phase (when we observe more of the western side). As a result, the spectral line contrast (linked to these gradients) will be weaker before the eclipse and stronger after the eclipse. b) panel, the fainter detection during pre-eclipse compared to post-eclipse observations could be due to the presence of clouds and the associated glory effect in the eastern hemisphere, a hint of this could be the hotspot offset detected in the optical CHEOPS and TESS light curves. }
	\label{hotspot}
\end{figure*}

\end{appendix}

\end{document}